# SOPE: A Spatial Order Preserving Encryption Model for *Multi*-dimensional Data

Eirini Molla
University of the Aegean
eirinim@aegean.gr

Theodoros Tzouramanis
University of the Aegean
ttzouram@aegean.gr

Stefanos Gritzalis
University of the Aegean
sgritz@aegean.gr

**Abstract**

Due to the increasing demand for cloud services and the threat of privacy invasion, the user is suggested to encrypt the data before it is outsourced to the remote server. The safe storage and efficient retrieval of d-dimensional data on an untrusted server has therefore crucial importance. The paper proposes a new encryption model which offers spatial order-preservation for d-dimensional data (SOPE model). The paper studies the operations for the construction of the encrypted database and suggests algorithms that exploit unique properties that this new model offers for the efficient execution of a whole range of well-known queries over the encrypted d-dimensional data. The new model utilizes well-known database indices, such as the $B^+$-tree and the R-tree, as backbone structures in their traditional form, as it suggests no modifications to them for loading the data and for the efficient execution of the supporting query algorithms. An extensive experimental study that is also presented in the paper indicates the effectiveness and practicability of the proposed encryption model for real-life d-dimensional data applications. This paper is an abridgment of a diploma thesis[1].

*Keywords*: applied cryptography; searchable encryption; order preserving encryption; seure queries; d-dimensional data;

## 1. Introduction

Cloud computing is a resources-delivery model, which provides services over the internet. Users can select and manage the available services, for example through a web browser, while the enterprise information system; its data and software are on servers at a remote location.

Cloud computing today plays a vital role in the implementation and execution of effective on-demand location-independent information systems' services in low cost. For this reason, cloud computing is becoming more and more popular while this popularity grows even more faster as more cloud services companies share valuable resources on the internet in cost effective ways.

In addition to the cloud services models that have been defined by NIST [1], the cloud offers also services such as the Database-as-a-Service (DaaS) which enables organizations to deploy new databases quickly, securely, and cheaply, thus offers them accelerated deployment, elastic capacity, great consolidation efficiency, and high availability in a very low overall operational cost and complexity.

Due to the increasing demand for more cloud services there is a growing threat of sensitive data security and privacy disruption by deploying several available network attacks [1, 2], such as SQL injection, cross site scripting, man in the middle, sniffing attacks, attacks against virtual machine hypervisors, *etc*. Therefore, the data needs to be encrypted in order to be protected, before stored in the cloud. By encrypting the data with an appropriate method, it is very difficult -if not impossible- for any potential attacker to gain access to the data in its non-encrypted form. However, the disadvantage of this solution is the communication and computational cost for executing the client's queries efficiently in the encrypted database.

The methods that offer sensitive data protection from untrustworthy cloud providers and malicious attackers and at the same time efficient execution of the client's queries can be grouped in several categories based on one of the following cryptographic primitives:

- *Searchable symmetric encryption*, in which every query over the encrypted database uses keywords and/or secure indices to retrieve the (encrypted) data results [3, 4, 5, 6, 7].
- *Fully-homomorphic encryption,* in which computations over encrypted data can be performed, based on a mathematical property called homomorphism that supports basic operations like addition and multiplication on ciphertexts [8, 9, 10].
- *Functional encryption*, in which the data owner encrypts the data using a public key and additionally s/he predefines access privileges for the rest of the users to access it. The users can then get secret keys from a trusted server and then decrypt parts of the encrypted data on the basis of their assigned privileges [11, 12, 13].

---

- *Oblivious Ram*, in which an interface between a program and the physical RAM can hide the user's access patterns from an adversary **[14, 15, 16, 17]**.
- *Property preserving encryption*, in which the encrypted data can sustain some selected properties of the original un-encrypted data **[18, 19, 20]**.

In the latter category, among other techniques, the Order Preserving Encryption (OPE) method is based on the principle that the order of two plaintexts $x$ and $y$ will hold also for their ciphertexts $Enc(x)$ and $Enc(y)$, i.e., $Enc(x) > Enc(y)$ iff $x > y$. The advantage of this method is that the order operations can be performed on the ciphertexts in the same way as on the plaintexts. Therefore query operations such as MIN, MAX and COUNT can be easily executed on the encrypted data as if they were on their original unencrypted form, while other operations such as SUM and AVG cannot function similarly since their output is not based on the order of the data. The disadvantage of the OPE method is that the cloud provider is aware of the order of the stored items. As a result, the provider can check out if a ciphertext is greater or smaller than any other ciphertext. Surely this is not a negligible disadvantage and under certain circumstances can hurt the security and privacy of the data. If for example the attacker is able to find the one-on-one correspondence between just one plaintext to its ciphertext, s/he may be able to approximately estimate the range of other ciphertexts through assumptions.

However there are applications in which revealing only the order of the data is not considered as a security leakage. As an example, an email storage box on a server can be considered, in which the data are encrypted with OPE on the date field or on the alphabetical order of the name of the sender. This way, given the OPE key, the email box owner can access the emails with a high degree of protection for the data while at the same time s/he can enjoy sorting and searching functionalities.

The example shows that as long as the application can endure that the OPE scheme will reveal information about the order of the items stored, the benefits on executing queries are significant since the technique allows comparison operations to be directly applied on the encrypted data, without the need to decrypting them.

The notion of *indistinguishability under the ordered chosen-plaintext attack* (IND-OCPA), **[20]** requires that an adversary with access to a set of ciphertexts will learn nothing except the order of the items. This guaranty is considered as the ideal security for OPE schemes for single-dimensional data and it has been shown that it is impossible for any OPE scheme to achieve IND-OCPA if the ciphertexts are immutable and the encryption is stateless.

**[21]** shows that IND-OCPA is impossible even for a stateful OPE scheme, however the property is achievable when the ciphertexts are mutable. The paper proposes also *the mutable order preserving encoding* (mOPE) model which uses an interactive protocol between the client and the untrusted server to help the latter managing the ciphertexts in the encrypted database.

This work proposes an OPE model for $d$-dimensional data that is influenced by the mOPE model for single-dimensional data proposed in **[21]**.

To the best of the authors' knowledge this work is the first to provide such a spatial OPE encryption for $d$-dimensional data. A whole range of traditional and other well-known spatial queries can be supported securely and efficiently within this new framework, as it is shown in the experimental study of the paper. In summary, the scientific contributions of this work are as follows:

- an interactive spatially-based mOPE scheme (SOPE) for d-dimensional data with a security guaranty that an adversary cannot distinguish between the encryptions of two sequences of objects as long as the sequences have the same spatial order relation,
- a bunch of algorithms for the secure and efficient processing of several well-known spatial queries, such as the point query, the range query, the $k$-nearest neighbours query, the static and dynamic skyline queries, *etc*.
- an extensive experimental evaluation study which illustrates the efficiency and practicability of the proposed scheme to support effectively high-demanding $d$-dimensional applications.

The rest of the paper is organized as follows. Firstly, the related work is discussed in **Section 2**. Threat models and some assumptions and notations that are relevant to the proposed work are discussed in **Section 3**. The proposed SOPE model for spatial order preservation in $d$-dimensional databases and a security analysis are presented in **Section 4**. The algorithms for processing an extensive set of queries in $d$-dimensional data using the new model are formally presented and discussed in **Section 5**. Extensive experimental results on real and synthetic data regarding the space and time performance of the proposed encryption framework are surveyed in **Section 6**. Finally, the conclusion of the research is provided in **Section 7**.

## 2. Related work

As it can be understood, OPE is a very popular model for single-dimensional database applications due to its ability to support efficient range query processing directly on the encrypted data without the need to decrypt them. The method was introduced in 2004 by Agrawal et al. **[19]** and a plethora of relative work has been introduced since then. **[19]** proposes the transformation of the plaintext database into a flat database such that the values are uniformly distributed. This flat database is then transformed into the cipher database such that the data values are distributed according to the targeted distribution. The transformation of the database is performed by splitting it in several buckets and by using linear interpolation inside every bucket. The drawback of this method is that it must take as input all the plaintexts in the database in advance which is not always practical in real-life applications. **[20]** shows that IND-OCPA is unachievable by any OPE scheme with stateless encryption and immutable ciphertexts. The authors propose an efficient OPE scheme on the basis of a sampling algorithm for the hypergeometric probability distribution. As IND-OCPA is unachievable for this scheme, they propose a security notion of a random order-preserving function (ROPF) and related primitives asking that an OPE scheme will look "as-random-as-possible" subject to the order preserving constraint. The encryption algorithm of the scheme in [20] behaves similarly to an algorithm that samples a ROPF from a specified domain on-the-fly (called "Lazy Sampling" in [22]). [23] improves the performance of the OPE scheme presented in [20] and [24, 25] show that the security definition of the ROPF inherently reveals at least half of the plaintext bits. Also, [26] experimentally show that this method has quite poor performance efficiency.

[27] proposes an OPE indexing scheme which indexes plaintexts by using simple linear expressions of the form $a * x + b + noise$, in which $x$ is the value for encryption, the coefficients $a$ and $b$ are kept secret (not known to the untrusted cloud server) and $noise$ is randomly sampled from some particular range, such that the order of the plaintexts is preserved. [28] proposes a nonlinear indexing scheme to address the vulnerability of linear indexing. A nonlinear indexing expression has the form $a * f(x) * x + b + noise$, where $f(x)$ is a function over $x$. However [21] discusses an attack that can break the schemes of [27, 28].

[29] weakens the security notion of IND-OCPA to IND-OLCPA (indistinguishability under ordered and local chosen-plaintext attack) requiring that the adversary can learn the ciphertexts only for 'nearby' values. The paper also extends the concept of OPE to generalized OPE (GOPE). Unlike OPE, the ciphertexts of GOPE may not be numbers, however using special comparison algorithms GOPE can still compare the encrypted data without needing to decrypt them. [30] proposes another generalization of OPE, called order-revealing encryption (ORE). In contrast to OPE in which the ordering of the underlying plaintexts is determined by numerically comparing the ciphertexts, in ORE scheme there is a publicly computable comparison function. The function takes two ciphertexts and outputs the numeric ordering of the underlying plaintexts. Although it provides IND-OCPA, the construction of the scheme relies on multi-linear maps and is therefore too inefficient in practice. An efficient ORE scheme is proposed by [31] however its drawback is that it leaks some information about the underlying plaintexts.

[32] presents a keyless IND-OCPA OPE scheme for outsourced data. The state of the algorithm plays the role of the key, *i.e.*, it is secret information. Differently from a key, the state of the algorithm is not pre-generated, but grows with the number of encryption operations. The size of the state of the encryption algorithm is the size of a dictionary of the database that maps the plaintext data to ciphertext which is sent to the server. If an update is required then potentially the entire client's dictionary and all the corresponding ciphertext values at the server-side might need to be updated. However, by keeping the dictionary at the client, the number of interactions between client and server for data updates are reduced. An important drawback of the method is its increased secure storage demands on the client-side.

[21] provides ideal security, which is IND-OCPA, by using a B$^+$-tree and an interactive protocol to provide OPE. Many other works have been proposed that rely on the OPE technology [19, 20, 24, 27, 28, 33, 34, 35, 36, 37, 38]. However, most of them provide weak security definitions by making assumptions about the possible attacks. [21], [34] and [32] are some excellent references providing comparisons regarding the security and efficiency of several previous works.

All the above-mentioned work has been proposed to handle single-dimensional data. As regards to multi-dimensional data, a solution to compute the range query is proposed in [39]. The method computes a secure indexing tag of the $d$-dimensional data by applying bucketization (*i.e.*, data splitting in buckets) which prevents the server from learning the exact plaintext values while it is still able to check if a record satisfies the $d$-dimensional range query predicate. The server returns a set of encrypted records and the client needs to perform some additional processes to select the records that satisfy the query. An analogous strategy with data grouping based on Voronoi diagrams is applied in [40] for the support of the nearest neighbour query.

[41] and [42] encrypt the database using similar data transformations and address the problem of nearest neighbour and skyline search respectively over the encrypted data. The drawback of the proposed query processing models is that their algorithms need to access every one tuple in the encrypted database to provide the answer to the query. [43] proposes another model to support the range query which however needs also to scan the whole database for executing the query. [44] focuses on location data and proposes linear coordinates transformations on every dimension, such as that in [27, 28] with the accompanied weaknesses of these schemes, to prevent the disclosure of the data while supporting efficiently range and $k$ nearest neighbor queries. The paper proposes also an R-tree-based solution with encrypted tree nodes (as encrypted buckets of records) which can be decrypted only by the client. In this solution most of the processes for data modifications and querying are performed by the client with high communication cost between the client and the server, while the server simply offers storage services for the encrypted R-tree.

Another transformation scheme [45] utilizes the Hilbert curve mapping [46] to transform spatial data points into one-dimensional values and then uses a one-dimensional OPE scheme to encrypt the data. The scheme supports the execution of range queries by decomposing them into a possibly large number of intervals, leading to high processing and communication costs. Also the evaluation of other spatial queries remains to be explored.

This paper extends the single-dimensional OPE methodology to the goal of supporting the encryption of $d$-dimensional data and the efficient execution of spatial queries on the encrypted database without the need to store any additional information but the secret key. The new scheme is inspired by [21] and utilizes an interactive communication protocol between the user and the server to offer IND-SOCPA, a security guaranty which is equivalent to IND-OCPA for multi-dimensional data. The cloud server is not assumed trusted and no intermediate authority such as a central trusted party is needed for any operation. The proposed model provides the user with the ability to access the database via any device with some minimum power and the server with the ability to implement its service using any existing spatial database management system (DBMS). The encryption procedure sustains the topology of the original data and to the best of the authors' knowledge this work is the first to provide such an encryption model offering spatial order preservation for *d*-dimensional data (SOPE model).

## 3. Preliminaries

### 3.1. Threat Models

It is assumed that the communication between the client and the server is performed without any intermediary entity (for example, a fully trusted authority) and that the client is capable to properly protect the secret key used for the encryption. It is also assumed that the cloud server to be honest-but-curious, whose goal might be to obtain full access to the plaintext of the encrypted stored data without altering any data that is communicated between the client and the server. The paper does not address data integrity and availability threats which can be handled by other mechanisms.

### 3.2. Assumptions and Notations

The proposal is based on the DaaS model. The two entities in the system are the legitimate client and the cloud server that interact between each other as the model is executed. The architecture considers the participation of one client although more clients can participate as well. The client owns the *d*-dimensional data and outsources them to the server in an encrypted form, wishing to not be revealed to any unauthorized entity. The client also aims to be able to search the data while protecting their confidentiality. The client is assumed to be capable to properly protect the secret key for the data decryption process. Finally, the client's device is assumed to have some minimum power, for example for being able to process the encryption and decryption processes or to perform simple calculations in order to refine the queries' results if needed.

The main burden of computation cost is assigned to the cloud service, which is this assumed to have ample storage space and power resources to store and query encrypted data through its sharing database services for the client. All data are assumed to be protected using existing symmetric or asymmetric data encryption schemes, which are not the focus of this paper, though symmetric encryption is encouraged.

Table 1 lists the most commonly used symbols in the paper.

| Symbol | Definition |
|---|---|
| OS | The original data space |
| ES | The 'encoded' data space that is produced by the proposed SOPE model |
| d | The number of data dimensions |
| P | A set of *d*-dimensional objects |
| P' | The encoded version of P |
| n | The cardinality of P |
| $p, e, r, p^1, p^2, .., p^k$ | *d*-dimensional data points |
| $p (p_1, p_2, …, p_d)$ | A *d*-dimensional point *p* with coordinates $p_1, p_2, …,$ and $p_d$ |
| $p' (p'_1, p'_2,…, p'_d)$ | The encoded version of point *p* |

Table 1: Symbols and Notations.

## 4. The proposed model

### 4.1. The new model's overview

The new model encrypts and stores *d*-dimensional objects on the basis of their spatial order. To achieve this it uses a separate B$^+$-tree for encoding the objects' coordinates in every dimension and an R-tree for finally storing the encoded objects. The system's simplified logical diagram is illustrated in **Figure 1**. A design goal is that the protocol of the proposed system should be simple enough to be implementable on top of existing clouds and DBMSs while at the same time it will operate with low communication and computational cost.

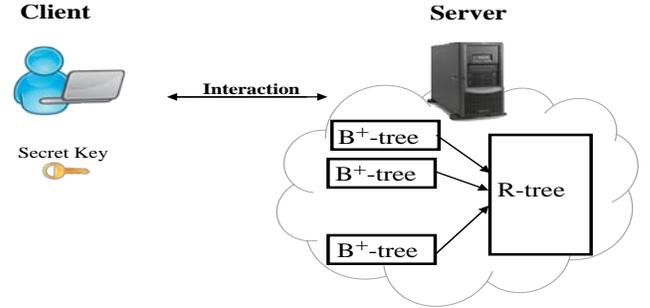

Figure 1: The simplified logical diagram of the proposed model.

### 4.2. Calculating the Encodings of the Encrypted Coordinates

For every dimension, a B$^+$-tree indexes the encrypted coordinate values of all the inserted database objects so far with the order of their corresponding plaintexts. This is achieved with the cooperation of the server and the user using a communication protocol that is based on the single-dimensional mOPE model proposed in **[21]** by adding some necessary modifications.

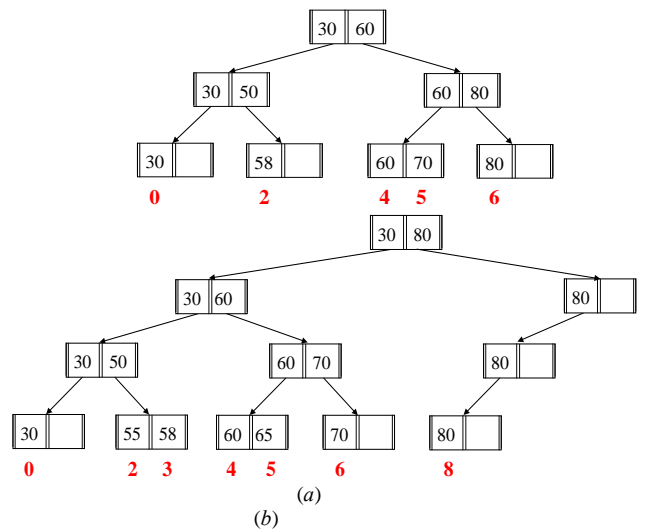

Figure 2: The evolution of a B$^+$-tree and the corresponding evolution of the encodings (in red colour under the leaves) of its stored keys.

Based on this strategy, an encoding is assigned to every single encrypted coordinate value; this encoding is related to the position of the coordinate value in the corresponding B$^+$-tree. To construct this encoding by following the model proposed in [21], $\lceil log_2B \rceil$ bits are needed for every level of the B$^+$-tree, where $B$ is the node's capacity, *i.e.*, $\lceil log_2B \rceil$ bits are needed to represent which pointer of every index node in the path from the root needs to be followed in order to find the coordinate value in the leaf level of the tree, and additionally $\lceil log_2B \rceil$ bits are needed to represent the position of this encrypted value in the hosting leaf. For example, assuming that the encrypted version of a coordinate value is 70, its encoding in the B$^+$-tree section of **Figure 2a** (2b) is 1-0-1 (0-1-1-0) in binary form, which can be converted to $1*2^2 + 0*2^1 + 1*2^0 = 5$ ($0*2^3 + 1*2^2 + 1*2^1 + 0*2^0 = 6$) in decimal form.

Analogous process is repeated for every dimension and thus the coordinates of every encrypted $d$-dimensional object are transformed to their corresponding encodings and the object is then inserted in the R-tree. This R-tree finally offers encryption with spatial order preservation since the order of the coordinate values of the stored objects does not change, thus preserving the corresponding spatial relations between the objects in the transformed space (*i.e.*, if for example an object $ob1$ is on the left side of an object $ob2$ before the encryption, this relation between the two objects is preserved in their encrypted version as well).

### 4.3. Data Insertion

This section surveys the object insertion process in the proposed SOPE model, with all the necessary communication interaction between the user and the server. For simplicity the discussion will focus on the insertion of a single-point object since the insertion of a multi-point object (e.g., a polygon) is a simply extension of this. The process is formally illustrated in **Algorithms 1** and **2**.

|    | Algorithm 1: Object Insertion ()<br>Input: a new point object $p(p_1, p_2, …, p_d)$.<br>Output: the updated B$^+$-trees with the OPE<br>  encodings for each dimension and the<br>  updated R-tree with the encoded objects. |
|----|---|
| 1: | Client:  informs the Server to begin the<br>  insertion process of the new point $p$; |
| 2: | FOR every dimension $i$ DO |
| 3: |   Server/Client: find in cooperation the<br>    position in the $i$-th B$^+$-tree in which<br>    the encrypted value of $p_i$ will be<br>    inserted, and find also as well its<br>    SOPE encoding $p'_i$; (see Algorithm 2) |
| 4: |   Server: inserts the encrypted value of<br>    $p_i$ coordinate in the $i$-th B$^+$-tree; |
| 5: |   Server: computes the range $[p^{1'}_i, p^{2'}_i]$ of<br>    the affected encodings in the B$^+$-tree<br>    that will be updated in the R-tree; |
| 6: |   Server: executes in the R-tree a range<br>    query $[p^{1'}_i, p^{2'}_i]$ on the $i$-axis to<br>    update any encoded $i$-coordinate of<br>    every object in this range to its<br>    newer value; |
| 7: | Server: Inserts the new point $p'(p'_1, p'_2,$<br>    $…, p'_d)$ in the R-tree; |
| 8: | END; |

Algorithm 1: The object insertion process in the proposed SOPE model.

In the begining of the process (**Algorithm 1**, Line 1) the client informs the server for the initiation of the insertion process. Since the coordinates of the object which the client wishes to insert are encrypted, the server cannot place the object in a spatial order with regard the other objects in the database. For this reason, for every dimension $i$ (**Algorithm 1**, Line 2), the server and the client interact (**Algorithm 1**, Line 3) in order to find the proper SOPE encoding $p_i'$ of the $i$-th coordinate $p_i$ of the object.

More specifically, as **Algorithm 2** shows in more detail, the server sends to the client the root node of the B$^+$-tree of the $i$-th dimension (**Algorithm 2**, Line 1). The client then decrypts the index keys in the root and compares them against the new key which s/he wants to insert into the tree. When the client finds the maximun decrypted index value that is smaller or equal than the key which will be inserted, s/he informs the server which pointer to follow in order to retrieve the correct child node in the next level of the tree (**Algorithm 2**, Line 3). The server receives this information and sends the corresponding child node to the client, which node -depending on the height of the tree- might be an index node or a leaf (**Algorithm 2**, Line 4). When the leaf level of the tree has been reached, the client (**Algorithm 2**, Line 5) informs the server for the position in which the new encrypted key needs to be inserted (if it does not already exist in the tree). The server then calculates the SOPE encoding $p_i'$ of the coordinate value $p_i$ based on the path from the root of the tree to the target leaf (**Algorithm 2**, Line 6). Details on the encoding calculation are provided in [21].

Then the server inserts the new encrypted coordinate value in the corresponding B$^+$-tree (**Algorithm 1**, Line 4). The server also computes the range $[p^{1'}_i, p^{2'}_i]$ -if exists- of affected encodings of pre-stored keys in the B$^+$-tree that need to be updated in the R-tree (**Algorithm 1**, Line 5) because the position that these keys or any of their ancestors had in the B$^+$-tree before this insertion, has now been changed.

In the next step the server updates in the R-tree the encoded $i$-coordinate of every spatial object that has value in the range $[p^{1'}_i, p^{2'}_i]$ it to its newer value (**Algorithm 1**, Line 6).

The same process (**Algorithm 1**, Lines 2 to 6) is repeated for every dimension and when the new item's coordinates encodings for all the space dimensions have been generated (based on the positions in which its encrypted coordinates have been stored in the corresponding B$^+$-trees), then the new multi-dimensional object is inserted into the R-tree (**Algorithm 1**, Line 7).

|    | Algorithm 2: Server/Client Interaction<br>  for Objects Insertion(), adapted from<br>  [21] |
|----|---|
| 1: | Server:  sends to the client the root node<br>  of the $i$-th B$^+$-tree; |
| 2: | FOR every non-leaf level of the tree |
| 3: |   Client:  decrypts the received node and<br>    compares its index keys with $p_i$.When<br>    s/he finds the proper index key,s/he |

```
            informs the server for its position;
4:      Server:  according to the position
            received by the client, s/he sends
            the corresponding child node to
            the Client;
5:      Client:decrypts the received leaf & sends
            to the Server the encrypted value of pᵢ
            and its position in the target leaf;
6:      Server: computes the SOPE encoding p'ᵢ of
            the encrypted value of pᵢ based on the
            path from the root of the tree to its
            position in the target leaf;
7:      END;
```

Algorithm 2: The interaction process for an object insertion in the proposed SOPE model.

As regards the encodings in the range $[p^{1\prime}_i, p^{2\prime}_i]$ of the *i*-th dimension whose values need to be updated in the R-tree, the following holds. If the insertion of the encrypted value of the new coordinate $p_i$ in the *i*-th B$^+$-tree has not caused any node split to the tree, then $p^{1\prime}_i$ and $p^{2\prime}_i$ are the old encodings of keys that lie both in the same leaf node with the newly inserted encrypted value of $p_i$. In this case, $p^{1\prime}_i$ is the old encoding of the key which is the next on the right side of the newly inserted encrypted value of $p_i$, and $p^{2\prime}_i$ is the old encoding of the key that is the right-most key in the same leaf, if both such keys exist. For example, in **Figure 2b**, after the insertion of key 55 in the B$^+$-tree of **Figure 2a**, the range $[p^{1\prime}_i, p^{2\prime}_i]$ is [2, 2], where 2 corresponds to the old encoding value of the key 58 whose position in the B$^+$-tree was affected from the insertion of the new key 55, and whose encoding value was changed from 2 to 3.

If, however, the insertion of the encrypted value of the new coordinate $p_i$ has caused the split of a tree node in at least one level of the B$^+$-tree, then $p^{1\prime}_i$ is the old encoding value of the next key on the right side of $p_i$ (not neseseraly being in the same leaf with $p_i$) and $p^{2\prime}_i$ is the right-most key in the right-most leaf an ancestor index key of which (*i.e.*, an ancestor index key in any possible level) changed its possition because of the insertion of $p_i$. For example, in **Figure 2b**, after the insertion of key 65 in the B$^+$-tree of **Figure 2a**, the tare phenomenon in which a node split in every level of the tree has appeared (with probability $(2/B)^h$, where $h$ is the level of the tree), therefore the range $[p^{1\prime}_i, p^{2\prime}_i]$ is $[5, +\infty]$, where 5 corresponds to the old encoding value of the key 70 which is the next key on the right side of 65 and $+\infty$ represents the right-most leaf-key in the tree since the encoding of every key whose 'greatest' ancestor index key is 80 (i.e., the index key 80 changed its position) needs to update its value because of this insertion. The result of the above strategy is that the encoding of zero keys in the best case (with probability between $1/B$ and $2/B$) and of all the allready-stored encrypted keys in the worst case (with probability $(1/n)_*(2/B)^h$) might need to be updated after the insertion of a new key.

*4.4. Security analysis*

The proposed encoding scheme is based on the single-dimensional mutable OPE model in **[21]** and extends it in the multi-dimensional space. The new model can mutate the encodings of the multi-dimensional objects in every dimension. Also, by preserving the order of the encodings in every dimension, the new multi-dimensional model manages to preserves the spatial order of the encrypted objects in the transformed space that is produced by these encodings.

Previous work on the field of OPE has shown that indistinguishability against chosen plaintext attack is unachievable by a practical OPE scheme. For this reason, a straightforward relaxation of this standard security notion for encryption in the single-dimensional domain is IND-OCPA **[20]**. For extending the IND-OCPA guarantee in the multi-dimensional domain, a corresponding security game is presented. The game is between the client and an adversary and it proceeds as follows:

1. The client chooses a random bit *b*.
2. The client and the adversary engage in a polynomial number of rounds of interaction in which the adversary is adaptive. At round *i*:
   a. The adversary sends to the client the sequences $P_i^0$ and $P_i^1$ of multi-dimensional objects.
   b. The client encrypts only the sequence $P_i^b$ of objects and by interaction with the server inserts these objects in the R-tree, with the adversary observing the result at the server-side.
3. The adversary outputs *b′*, its guess for *b*.

The adversary will win the game if the guess is correct (i.e., if $b = b\prime$), while the sequences $P_i^0$ and $P_i^1$ are comprised by objects that have the same spatial order relation among each other in both sequences of objects.

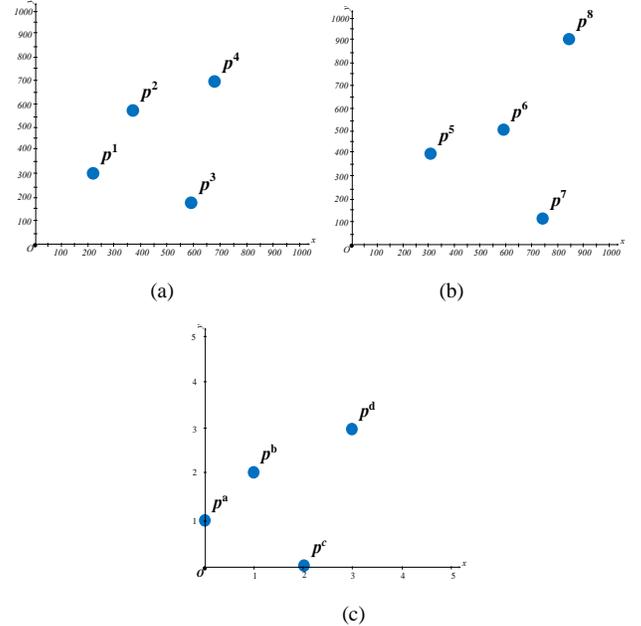

Figure 3: (a) and (b) two sequences of objects in the OS with the same spatial order between their objects, and (c) the encoded version of both these objects sequences in the ES.

An example of such sequences of objects that have the same spatial order is depicted in **Figure 3**. **Figures 3.(a)** and **3.(b)** illustrate the sequences $\{p^1, p^2, p^3, p^4\}$ and $\{p^5, p^6, p^7, p^8\}$ of 2-dimensional points in the original unencrypted space (OS). The object $p^2$ ($p^6$) is in the north-east side of object $p^1$ ($p^5$), the object $p^3$ ($p^7$) is in the south-east side of object $p^2$ ($p^6$), *etc.*

**Figure 3.(c)** illustrates the encoded version {$p^a$, $p^b$, $p^c$, $p^d$} of one of these sequences of points in the encoded space (ES) that is provided by the SOPE model. The adversary will win the game if s/he can guess correctly which of the two sequences was encrypted. Let *win* be the random variable indicating the success of the adversary in the above game.

***Definition* 1 (*IND-SOCPA – indistinguishability under a spatial-ordered chosen-plaintext attack*):** A SOPE scheme is IND-SOCPA secure if for all p.p.t. adversaries: $Pr[win] \leq \frac{1}{2} + negl(.)$, where $negl(.)$ is a negligible function **[47]**.

***Theorem.*** The proposed SOPE scheme is IND-SOCPA secure.

***Proof.*** The proof is a simple extension of the proof of the corresponding theorem that the backbone mOPE scheme is IND-OCPA secure, which provided in **[21]**. For this reason a short proof intuition is only provided in this paper. Consider any adversary and any two sequences $P^0$ and $P^1$ of $d$-dimensional objects that the adversary asks for in the security game. The view of the adversary consists of the information the server receives in the security game.

Assuming that the data encryption scheme used by the client (which is not the focus of this paper) produces computationally indistinguishable from random values that have the same pattern of repetitions (e.g., produced by a random oracle), the discussion focuses on the role of the SOPE scheme in the encoding process. For this goal, we proceed inductively in the number of $d$-dimensional objects to be encrypted. Assuming that after the insertion of $i$ objects in which the adversary obtains the same information in both cases of sequences $P^0$ and $P^1$, it will be shown that the information the adversary will obtain after the $i+1$-th insertion will be equally the same again. Supposing that $u$ is the $i+1$-th object in the $P^0$ sequence and $v$ is the $i+1$-th object in the $P^1$ sequence, in both cases the client and the server in cooperation will execute **Algorithm 1** to insert the object in the R-tree. For every one of the $d$ coordinate values of either object $u$ or $v$, since both $P^0$ and $P^1$ sequences provide the same order in every dimension, the path down on every B$^+$-tree taken by the algorithm is the same. Either or not the encrypted coordinate value will exist in the corresponding B$^+$-tree, the only information the client gives to the server is which child tree nodes to take in this path, which is also the same for both cases. And also in both cases the insertion into the R-tree will be exactly in the same position in the tree since the encodings of both $u$ and $v$ on all dimensions will be exactly the same. Therefore, the adversary receives the same information in both cases, and hence s/he cannot distinguish between them. □

In the example of **Figure 3**, the encoded result in **Figure 3.(c)** that is provided by the proposed SOPE model might have been produced by any one of the two sequences of points in **Figures 3.a** or **3.b** with the same probability ½ (for example the encoded version of both the points $p^2$ of the first sequence and $p^6$ of the second sequence is the point $p^b$(1, 2) in the encoded space).

## 5. Query processing

In this section, algorithms for the efficient execution of a bunch of well-known queries for $d$-dimensional data are presented, which are: the point query, the range query, the (static) skyline query, the dynamic skyline query, the $k$-nearest neighbours query, the constrained $k$-nearest neighbours query, the constrained skyline query, the reverse $k$-nearest neighbour query, the constrained skyline query, and the continuous nearest neighbour query. An analysis on the skyline query family can be found in **[48]**. For every query, an example is presented, as well as a formal description of the proposed algorithm to efficiently support it, using a step-by-step pseudo-code illustration, and a theoretical proof of its correctness and efficiency. The 2-dimensional dataset $P$ which is used throughout all the examples that follow is illustrated in Table 2.

| object | X | Y | X' | Y' | object | X | Y | X' | Y' |
|---|---|---|---|---|---|---|---|---|---|
| $p^1$ | 100 | 100 | 1 | 1 | $p^{15}$ | 50 | 950 | 0 | 20 |
| $p^2$ | 250 | 250 | 4 | 4 | $p^{16}$ | 900 | 750 | 17 | 16 |
| $p^3$ | 600 | 600 | 11 | 13 | $p^{17}$ | 950 | 950 | 18 | 20 |
| $p^4$ | 300 | 400 | 5 | 7 | $p^{18}$ | 400 | 50 | 7 | 0 |
| $p^5$ | 450 | 450 | 8 | 8 | $p^{19}$ | 750 | 250 | 14 | 4 |
| $p^6$ | 100 | 700 | 1 | 15 | $p^{20}$ | 850 | 150 | 16 | 2 |
| $p^7$ | 300 | 480 | 5 | 9 | $p^{21}$ | 150 | 650 | 2 | 14 |
| $p^8$ | 500 | 900 | 9 | 19 | $p^{22}$ | 100 | 200 | 1 | 3 |
| $p^9$ | 800 | 550 | 15 | 12 | $p^{23}$ | 550 | 100 | 10 | 1 |
| $p^{10}$ | 350 | 850 | 6 | 18 | $p^{24}$ | 700 | 510 | 13 | 11 |
| $p^{11}$ | 200 | 300 | 3 | 5 | $P^{25}$ | 700 | 800 | 13 | 17 |
| $p^{12}$ | 650 | 150 | 12 | 2 | $p^{26}$ | 700 | 350 | 13 | 6 |
| $p^{13}$ | 950 | 900 | 18 | 19 | $P^{27}$ | 100 | 350 | 1 | 6 |
| $p^{14}$ | 600 | 300 | 11 | 5 | $p^{28}$ | 100 | 500 | 1 | 10 |

Table 2: A dataset $P$ of twenty eight 2-dimensional points with coordinates <$X$, $Y$> and SOPE encoding values <$X'$, $Y'$>.

### 5.1. The Point Query

Given a dataset $P$ of $d$-dimensional points and a reference query point $q$, the point query finds if $q$ exists in the dataset. The algorithm for processing the point query in the proposed SOPE model is illustrated in **Algorithm 3**.

```
    Algorithm 3: The Point Query()
    Input:   a reference point q(q₁, q₂, …, qd).
    Output:  TRUE if q appears in the dataset
      and FALSE if it does not.
1:  Client: informs the Server to begin the
      execution of the point query with  regard to
      the given query predicate q;
2:  FOR every dimension i DO
3:     Server: sends to the client the root
          node of the i-th B⁺-tree;
4:     FOR every non-leaf level of the tree
5:        Client: decrypts the received node
             and compares its index keys
             with qᵢ. When s/he finds the
             proper index key, s/he informs
             the server for its position;
6:        Server: according to the position
             received by the client, s/he
             sends the corresponding child
             node to the Client;
7:     Client: decrypts the received leaf;
8:     IF qᵢ is found in the leaf THEN
```

```
 9:         Client: sends to the Server the
               position of q_i in the leaf;
10:         Server: computes the SOPE encoding
               q'_i of q_i based on the path from
               the root of the tree to the leaf;
11:       ELSE
12:         return FALSE;
13:   Server: executes a traditional point
           query to the R-tree with regard to the
           encoded query predicate q' and returns
           FALSE if q does not appear in the
           dataset;
14:   return TRUE;
```

Algorithm 3: The point query process in the proposed SOPE model.

In the begining of the process, after in Line 1 of the algorithm the client informs the server for the initiation of the execution of the point query with regard to the given query predicate $q(q_1, q_2, …, q_d)$, for every dimension $i$ (Line 2) the server and the user interact (Lines 3-12) using a communication protocol similar to the one appearing in **Algorithm 2** for the object insertion process, in order to find if $q_i$ exists in the corresponding $i$-th B$^+$-tree. If $q_i$ is not found the process is terminated (Line 12) since it is evident that non object in the dataset will have coordinate on the $i$-th axis that is equal to $q_i$. Otherwise the encoding $q'_i$ of $q_i$ is calculated and when this calculation has been succesfully performed for all dimensions, a traditional point query is executed in the R-tree with regard to the encoded query predicate $q'(q'_1, q'_2, …, q'_d)$ to discover if $q$ truly appears in the dataset.

## 5.2. The Range query

Given a dataset $P$ of $d$-dimensional points and a reference query hyper rectangle $q$ with diagonal vertices $qa$ and $qb$, the range query retrieves the points that lie in the given rectangle. The algorithm for processing the range query in the proposed SOPE model is illustrated in **Algorithm 4**.

```
     Algorithm 4: The Range Query()
     Input: a reference hyper rectangle q with
         diagonal vertices qa(qa_1, qa_2, …, qa_d),
         and qb(qb_1, qb_2, …, qb_d).
     Output: the data points that lie into q.
1:   Client: informs the Server to begin the
         execution of the range query with regard to
         the given hyper rectangle q;
2:   Client/Server: calculate in cooperation
         the encoded version q'a and q'b of the
         points qa and qb, correspondingly;
3:   Server: executes the traditional Range
         Query to the R-tree with regard to q'
         with diagonal vertices q'a and q'b;
4:   Server: FOR every data point in the
         results, the B^+-tree for every
         dimension is traversed to collect the
         ciphertext of its coordinates and sends
         the point to the Client;
5:   Client: decrypts the results;
6:   END;
```

Algorithm 4: The range query process in the proposed SOPE model.

In the beginning of the process (Line 2), the client and the server interact with each other in order to calculate the encodings $q'a$ and $q'b$ of the diagonal vertices $qa$ and $qb$ of the given hyper rectangle $q$, as if $q'a$ and $q'b$ should be inserted in the dataset, using a communication protocol similar to the one appearing in Algorithm 2 for the object insertion process. Then a traditional range query is executed in the R-tree with regard to the encoded hyper rectangle $q'$ with diagonal vertices $q'a$ and $q'b$ (Line 3), to retrieve all point objects in $P$ that appear to be inside the requested region. Then for every data point in the results, the server traverses the corresponding B$^+$-tree in every dimension to collect the ciphertext of the coordinates of the point (Line 4) and then sends the point to the client for descryption. The following proposition validates the correctness of the algorithm.

***Proposition* 1***:*** The execution of the range query over the original (un-encrypted) as well as over the encoded dataset using the proposed SOPE model produces the same results.

***Proof:*** Let us assume a $d$-dimensional hyper rectangle $q$ with diagonal vertices $[qa, qb]$ and its encoded version $q'$ using the SOPE model with diagonal vertices $[q'a, q'b]$, where $q'a$ and $q'b$ are correspondingly the SOPE encodings of $qa$ and $qb$. Let us also assume that the execution of the range query $[qa, qb]//[q'a, q'b]$ over the original//encoded dataset $P//P'$ will produce as result the set of points $ORQ//ERQ$. Let us assume now a point $p \in ORQ$, then $\forall\ i \in \{1, …, d\}: qa_i \le p_i \le qb_i$ and since the SOPE model does not change the order of these values on the $i$-th axis, then $\forall\ i \in \{1, …, d\}: q'a_i \le p'_i \le q'b_i$, where $q'a_i, p'_i$ and $q'b_i$ are correspondingly the OPE encodings of $qa_i, p_i$ and $qb_i$. Therefore, $p \in ERQ$, thus $ORQ \subseteq ERQ$. Similarly it can be shown that $ERQ \subseteq ORQ$, therefore, finally, $ORQ \equiv ERQ$. □

A range query example in the 2-dimensional original space OS as well as in the corresponding encoded space ES is illustrated in **Figure 4**. As the figure shows, the range query provides the same results in both domains. As it can also be understood, the SOPE model does not preserve the distance between points, therefore, point $p^7$ in the OS is closer to the upper side of the query rectangle, than it is in the ES.

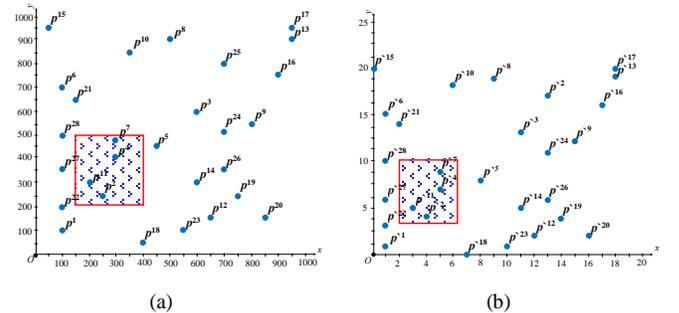

Figure 4: The range query: (a) in the OS, and (b) in the ES.

## 5.3. The Skyline Query

The skyline operator is important for several applications involving multi-criteria decision making. Given a dataset $P$ of $d$-dimensional points, the operator returns all data points $p$ that are not dominated by another point *r* in the dataset. For simplicity, we assume that skylines are computed with respect to *min* conditions on all dimensions. Using the *min* condition, a point *p* dominates another point *r* if and only if the coordinate of *p* on any axis is not larger than the corresponding coordinate of *r*. Informally, this implies that *p* is preferable (or 'more interesting') to *r* according to any preference (scoring) function which is monotone on all attributes. The algorithm for processing the skyline query using the proposed SOPE model is illustrated in **Algorithm 5**.

|   | Algorithm 5: The Skyline Query()<br>Output: the data points that belong into the dataset's skyline. |
|---|---|
| 1: | Client: informs the Server to begin the execution of the skyline query; |
| 2: | Server: executes the traditional skyline query in the R-tree; |
| 3: | Server: FOR every data point in the results, the B⁺-tree for every dimension is traversed to collect the ciphertext of its coordinates and sends the point to the Client; |
| 4: | Client: decrypts the results; |
| 5: | END; |

Algorithm 5: The skyline query process in the proposed SOPE model.

The algorithm shows that the skyline query can be processed directly on the encoded data by executing any traditional skyline query processing algorithm for indexed data (for example the one introduced in **[49]** and its results will be valid also in the original unencrypted domain. This can be verified by the following proposition.

***Proposition 2:*** The execution of the skyline query over the original (un-encrypted) as well as over the encoded dataset using the proposed SOPE model produces the same results.

***Proof***: Let us assume that the execution of the skyline query over the original//encoded *d*-dimensional dataset *P*//*P′* will produce as a result the set of points *OSQ*//*ESQ*. Let us also assume a point $p \in OSQ$ and its corresponding encoded point $p' \in P'$ and that $p' \notin ESQ$. Therefore, there is a point $r' \in ESQ$ such that $\forall i \in \{1, ..., d\}: r'_i \leq p'_i$, and since the SOPE encoding does not change the order of $r_i$ and $p_i$, we have that $\forall i \in \{1, ..., d\}: r_i \leq p_i$, therefore $p \notin OSQ$, which is a contradiction to the initial hypothesis that $p \in OSQ$. Thus $p' \in ESQ$, and thus $OSQ \subseteq ESQ$. Similarly it can be shown that $ESQ \subseteq OSQ$, therefore, finally, $OSQ \equiv ESQ$. □

A skyline query example in the 2-dimensional original space OS as well as in the corresponding encoded space ES is illustrated in **Figure 5**. As the figure shows, the skyline query provides the same results in both domains.

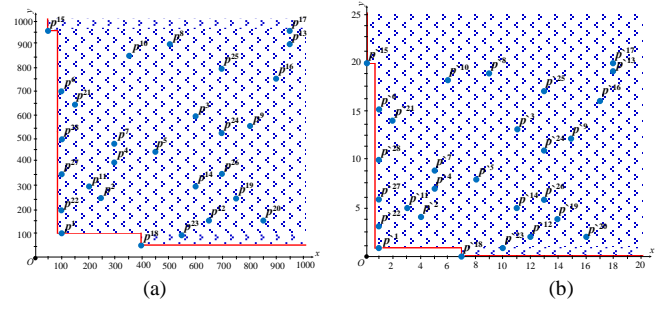

Figure 5: The skyline query: (a) in the OS, and (b) in the ES.

### 5.4. The Global Skyline Query

Given a dataset *P* of *d*-dimensional points and a reference query point *q*, the global skyline query with regard to *q* retrieves the set GSL(*q*) of those data points which are not globally dominated by another point in the dataset according to *q*. The formal definition of global domination follows.

***Definition 2 (Global domination* [50])**: a point $p \in P$ globally dominates a point $r \in P$ with regard to the query point *q* iff:

1. $\forall i \in \{1, ..., d\}: (p_i - q_i)(r_i - q_i) > 0$,
2. $\forall i \in \{1, ..., d\}: |p_i - q_i| \leq |r_i - q_i|$ and
3. $\exists j \in \{1, ..., d\}: |p_j - q_j| < |r_j - q_j|$.

The algorithm for processing the global skyline query using the proposed SOPE model is illustrated in **Algorithm 6**. The Line 3 of the algorithm uses the corresponding global skyline query algorithm for processing un-encrypted data proposed in **[50]**.

|   | Algorithm 6: The Global Skyline Query()<br>Input: a reference point $q(q_1, q_2, ..., q_d)$.<br>Output: the data points that belong to the global skyline of *q*. |
|---|---|
| 1: | Client: informs the Server to begin the execution of the global skyline query with regard to the query predicate *q*; |
| 2: | Client/Server: calculate in cooperation the encoded version *q′* of *q*; |
| 3: | Server: executes the traditional global skyline query with regard to *q′*, using the R-tree (see [50]); |
| 4: | Server: FOR every data point in the results, the B⁺-tree for every dimension is traversed to collect the ciphertext of its coordinates and sends the point to the Client; |
| 5: | Client: decrypts the results; |
| 6: | END; |

Algorithm 6: The global skyline query process in the proposed SOPE model.

The above **Algorithm 6** in Line 3 shows that the global skyline query can be processed directly on the encoded data by executing the traditional global skyline query processing algorithm proposed in **[50]** and its results will be valid in the original unencrypted domain as well. This can be verified by the following proposition.

***Proposition 3:*** The execution of the global skyline query over the original as well as over the encoded dataset using the proposed SOPE model produces the same results.

***Proof:*** Let us assume that the execution of the global skyline query over the original//encoded $d$-dimensional dataset $P$//$P'$ will produce as the result the set of points $OSQ$//$ESQ$. Let us also assume a global skyline point $p \in OSQ$ for which we have that $\forall\ i \in \{1, …, d\}$: $q_i \leq p_i$, and that for its corresponding encoded point $p' \in P'$ we have $p' \notin ESQ$. Therefore, there is a point $r' \in ESQ$ such that $\forall\ i \in \{1, …, d\}$: $r_i' \leq p_i'$, and since the SOPE encoding does not change the order of $r_i$ and $p_i$, we have that $\forall\ i \in \{1, …, d\}$: $r_i \leq p_i$, therefore $p \notin OSQ$, which is a contradiction to the initial hypothesis that $p \in OSQ$. Thus $p' \in ESQ$, and thus $OSQ \subseteq ESQ$. Similarly we can show that $ESQ \subseteq OSQ$, therefore, finally, $\forall p \in OSQ$ for which we have that $\forall\ i \in \{1, …, d\}$: $q_i \leq p_i$ we have that $OSQ \equiv ESQ$.

Similarly we can show that $OSQ \equiv ESQ$ for all the hyper quadrants of the data space for which $q$ is the origin of the axes. **Figure 6** illustrates the four quadrants of the space that are created having $q'$ as the origin of the axes when we study this problem in two dimensions. □

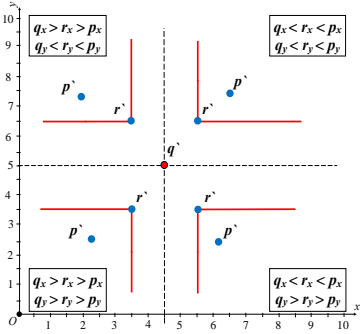

Figure 6: Global skyline query processing with regard to a point $q'$ in the 2-dimensional ES, and the four possible spatial relations between a global skyline point $r'$ and another point $p'$ under examination.

A global skyline query example with regard to a reference point $q$ in the 2-dimensional OS as well as in the corresponding ES is illustrated in **Figure 7**. As the figure shows, the query provides the same results in both domains.

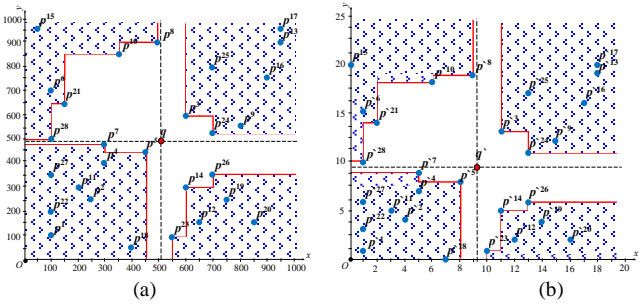

Figure 7: The global skyline query: (a) in the OS, and (b) in the ES.

### 5.5. The Dynamic Skyline Query

Given a dataset $P$ of $d$-dimensional points and a reference query point $q$, the dynamic skyline query specifies a transformed $d$-dimensional space which is built based on the original space. Every point $p$ in $P$ is mapped to another point with coordinates $f_1(p),…,f_d(p)$, where $f_i$ is a one-dimensional function $\forall\ i \in \{1, …, d\}$. The dynamic skyline of $P$ with respect to the functions $f_1, …, f_d$ returns the ordinary (static) skyline of the transformed $d$-dimensional space. For simplicity, we assume that for every point $p$ in $P$ we have: $\forall\ i \in \{1, …, d\}$: $f_i(p) = |q_i - p_i|$, i.e., the function $f_i$ simply refers to the absolute distance of every point $p$ to the query point $q$ in the $i$-th dimension. The following well-known definitions clear the related terms that are used to this research.

***Definition 3 (Dynamic Skyline):*** Given a data set $P$ and a query point $q$, the dynamic skyline query with regard to $q$ retrieves the set $DSL(q)$ of all data points in $P$ that are not dynamically dominated with regard to $q$.

***Definition 4 (Dynamic Domination):*** A data point $p \in P$ dynamically dominates another data point $r \in P$ with regard to $q$ iff:
1. $\forall\ i \in \{1, ..., d\}$: $|q_i - p_i| \leq |q_i - r_i|$, and
2. $\exists\ j \in \{1, ..., d\}$: $|q_j - p_j| < |q_j - r_j|$.

The example in **Figure 8** shows that the execution of the dynamic skyline query with regard to a query point $q$ on the original (un-encrypted) dataset does not produce the same results with the execution of the same query on the encoded dataset using the proposed SOPE model. For instance, although the data point $p^{24}$ belongs to dynamic skyline of $q$ in the OS, its corresponding encoded data point $p'^{24}$ does not belong to dynamic skyline of the encoded version $q'$ of $q$ in the ES because the data point $p''^{7}$ dominates it.

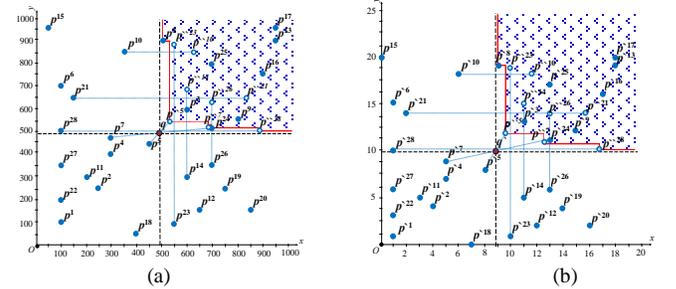

Figure 8: The dynamic skyline query: (a) in the OS, and (b) in the ES.

To overcome this problem, the global skyline query can be used to support the calculation of the dynamic skyline query in the SOPE model. This is supported by the following proposition.

***Proposition 4:*** Let $P$ be a $d$-dimensional dataset, $q$ be the query point, $GSL(q)$ be the set of global skyline points of $P$ with regard to $q$ and $DSL(q)$ be the set of dynamic skyline points of $P$ with regard to $q$. It holds that $DSL(q) \subseteq GSL(q)$.

***Proof:*** Let us suppose that we have a point $p \in DSL(q)$ for which $p \notin GSL(q)$. Therefore, there is a point $r \in GSL(q)$ that globally dominates $p$. Therefore, since based on Definition 2 we have that $\forall\ i \in \{1, ..., d\}$: $|p_i - q_i| \geq |r_i - q_i|$ and $\exists\ j \in \{1, ..., d\}$: $|p_j - q_j| > |r_j - q_j|$, it follows that based on Definition 4 we have that $p \notin DSL(q)$, which is a contradiction to our hypothesis. Therefore $DSL(q) \subseteq GSL(q)$. □

On the basis of Propositions 4 and 3, the execution of a dynamic skyline with regard to a query point $q$ in the OS can be translated to an execution of a global skyline in the ES, that will be followed by a final refinement step to select the global skyline points with regard to $q$ that are also dynamic skyline points with regard to $q$. The proposed algorithm for processing the dynamic skyline in the SOPE model is therefore as illustrated in **Algorithm 7**.

|   | Algorithm 7: The Dynamic Skyline Query() Input: a reference point $q(q_1, q_2, …, q_d)$. Output: the data points that belong to the dynamic skyline of $q$. |
|---|---|
| 1: | Client/Server: execute in cooperation the Algorithm 6; |
| 2: | Client: executes the traditional dynamic skyline query to the decrypted results in order to select the dynamic skyline set with regard to $q$; |
| 3: | END; |

Algorithm 7: The dynamic skyline query process in the proposed SOPE model.

Therefore, as **Figure 9** shows, in the case of the dynamic skyline query the Server in the ES will return to the Client the global skyline set with regard to the query predicate $q$ (**Figure 9.(a)**). Then the Client will decrypt the results and will provide the final refinement step to execute on this selected subset of the initial dataset any well-known algorithm for processing the dynamic skyline -with or without using an index- to retrieve the dynamic skyline (**Figure 9.(b)**).

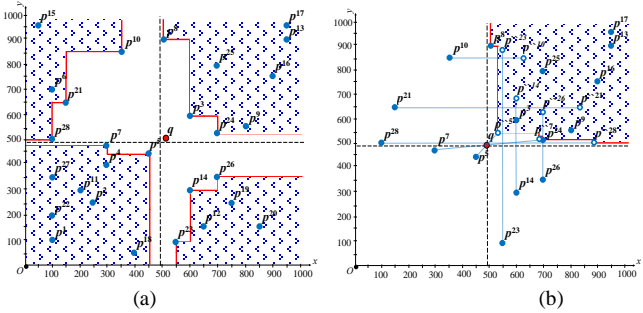

Figure 9: Processing the dynamic skyline query: (a) in the first step the Server will calculate the global skyline set in the ES, and (b) in the second step the Client in the OS will calculate the dynamic skyline of the results of the global skyline received by the Server.

*5.6. The k-Nearest Neighbour Query*

Given a dataset $P$ of $d$-dimensional points and a query reference point $q$, the $k$-nearest neighbour ($k$NN) query retrieves the $k$ nearest neighbour data points of $q$ according to a specified distance metric (Euclidean distance is assumed as the distance metric in this study).

However, as it can be understood, the proposed SOPE model maintains only the order of the data in every axis, not the order of the distance from the origin of the axes or the order of the distances between the data points. Therefore, the execution of a nearest neighbour query will not provide necessarily the same results in the OS and in the ES. **Figure 10** provides such an example, in which the nearest neighbour query with regard to a reference point $q$ in the OS will retrieve the data point $p^3$ while the same query with regard to the corresponding encoded point $q'$ in the ES will retrieve the data point $p'^{24}$. This happens because in the OS the data point $p^3$ is closer to $q$ than $p^{24}$ and also than any other point, however the opposite holds in the ES.

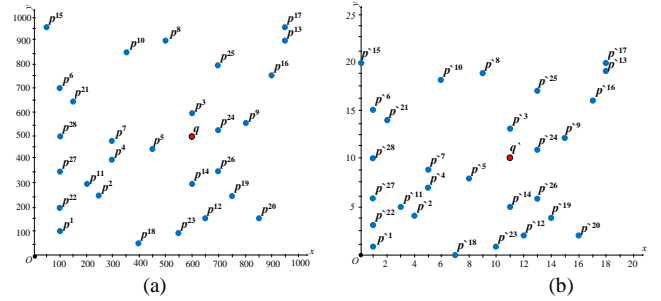

Figure 10: The nearest neighbour query with regard to a reference point $q$: (a) in the OS, and (b) in the ES.

To overcome this problem, the $k$-global skyline query (*i.e.*, a generalization of the well-known global skyline query for $k \geq 1$) can be used to support the calculation of the $k$NN query in the SOPE model. This is supported by the following observation.

***Observation* 1*:*** The $k$-global skyline set $k$-GSL($q$) of a dataset $P$ with regard to a query point $q$ includes the $k$ nearest neighbour points of $P$ to $q$.

***Proof*:** In the case of $k = 1$, in Proposition 4 it has been proved that the set 1-GSL($q$) of the first order of global skyline of $P$ with regard to a reference point $q$ is a superset of the dynamic skyline set with regard to $q$. Also, in **[51]** it has been proved that the dynamic skyline set of $P$ with regard to $q$ includes the nearest neighbour point $p$ of $P$ to $q$. Therefore the observation truly holds for $k = 1$.

In the case of $k = 2$, if we remove from $P$ the set of the first order of global skyline of $P$, then using **[50]** we can find the set of the second order of global skyline of $P$ with regard to $q$, in which we will also have the nearest neighbour point $r$ of the set "$P$ – 1-GSL($q$)" to $q$. It is obvious now that the set 1-GSL($q$) $\cup$ $\{r\}$ includes the 2 nearest neighbour points of $P$ to $q$, therefore the observation truly holds for $k = 2$. Using the same arguments it can be proved by induction that the observation holds $\forall$ $k \in N$. □

The following proposition is an extension of Proposition 3 for $k \geq 1$ and its proof is also analogous to that of that previous proposition.

***Proposition* 5*:*** The execution of the $k$-global skyline query over the original as well as over the encoded dataset using the proposed SOPE model produces the same results.

**Algorithm 8** processes the $k$NN query in the proposed SOPE model. Therefore, if the client wishes to execute the $k$NN query with regard to a query point $q$, then in Line 3 of the algorithm the server will execute the corresponding $k$-global skyline query with regard to the encoded point $q'$ (the supporting **Algorithm 12** for processing of the $k$-global skyline query is presented in the Appendix), and after collecting the data points in the results and the ciphertext of their coordinates (Line 4), the server will send the results to the client. The client in Line 5 will then decrypt the results to compare the distance from $q$ of every point in the result set, in order to find the $k$ points with the minimum distance, *i.e.*, the $k$ nearest neighbour points to $q$.

```
    Algorithm 8: The kNN Query()
    Input:   a reference point q(q₁, q₂, …, q_d)
         and the requested value of k.
    Output:  the kNN data points of q
1:  Client: informs the Server to begin the
        execution of the kNN query   with regard
        to the query predicate q;
2:  Client/Server: calculate in cooperation
        the encoded version q' of q;
3:  Server: executes the k-global skyline
        query with regard to q', using the R-tree
        (see Algorithm 12);
4:  Server: FOR every data point in the
        results, the B⁺-tree for every
        dimension is traversed to collect the
        ciphertext of its coordinates and sends
        the point to the Client;
5:  Client: decrypts the results & compares their
        distances from q to find the kNN of q;
6:  END;
```

Algorithm 8: The $k$NN query process in the proposed SOPE model.

A $k$-global skyline query example with regard to a reference point $q$ in the 2-dimensional OS and the corresponding ES for $k$=1 // $k$=2 is illustrated in **Figure 7** // **Figure 11**. As the figures show, the query provides the same results in both domains. An interesting observation is that as **Figure 11** shows, for $k = 2$ the two nearest points to $q$ in the OS are the points $p^3$ and $p^{24}$ which both happen to be part of the 1st order of the global skyline with regard to $q$.

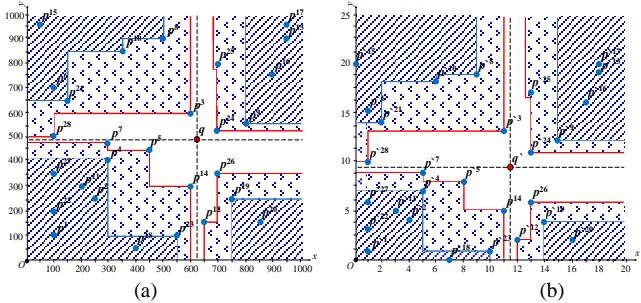

Figure 11: The 2-global skyline query: (a) in the OS, and (b) in the ES.

## 5.7. The Constrained k-Nearest Neighbour Query

Given a dataset $P$ of $d$-dimensional points, a reference query point $q$ and a hyper rectangle $r$, the constrained $k$-nearest neighbour (constrained $k$NN) query retrieves the $k$NN data points of $q$ that lie in the specified region $r$. Therefore, the query targets nearest neighbour points in a region bounded by certain spatial conditions **[52]**.

As **Algorithm 9** shows, a technique to execute the constrained kNN on the indexed encoded data stored in the ES is to execute a $k$-global skyline query on these data that belong in the desired spatial region (Line 3). After collecting the data points in the results and the ciphertext of their coordinates (Line 4), the server will send the final result set to the client for decryption and calculation of the $k$ nearest neighbours of $q$ in the constrained region (Line 5). Conceptually, as in the case of the $k$NN query, the calculation of the constrained $k$NN query using the new SOPE model has two phases. The first phase is a filtering phase with the execution of a global skyline query in a set of data points that belong in a desired spatial region in the ES. The second phase is a refinement phase with the execution of a traditional $k$NN query in the OS, using the small subset of points that has been selected by the filtering phase.

```
    Algorithm 9: The Constrained kNN Query()
    Input: a reference d-dimensional hyper re-
        ctangle r with diagonal vertices ra(ra₁,
        ra₂, …, ra_d), and rb(rb₁, rb₂, …, rb_d),
        a query point q(q₁, q₂, …, q_d) and the
    requested value of k.
    Output:  the kNN data points of q in the desired
    region r.
1:  Client: informs the Server to begin the
        execution of the constrained kNN query with
    regard to the hyper rectangle r and
        the query point q;
2:  Client/Server: calculate in cooperation
        the encoded version r' (i.e., the
        encoded points r'a and r'b) of the
        rectangle r (i.e., of the points ra and
        rb, correspondingly) and the encoded
        version q' of q;
3:  Server: executes the constrained k-global
        skyline query with regard to r' and q',
        using the R-tree;
4:  Server: FOR every data point in the
        results, the B⁺-tree for every
        dimension is traversed to collect the
        ciphertext of its coordinates and sends
        the point to the Client;
5:  Client: decrypts the results & compares their
        distances from q to find the kNN of q;
6:  END;
```

Algorithm 9: The constrained $k$NN query process in the proposed SOPE model.

The supporting **Algorithm 14** for processing the constrained $k$-global skyline query is presented in the Appendix.

Two constrained $k$NN query examples with regard to a reference point $q$ in the 2-dimensional OS and the corresponding ES with $k$=1 are illustrated in **Figures 12** and **13**. In the first figure the query point $q$ is located inside the constrained region while in the second figure the query point is located outside the constrained region.

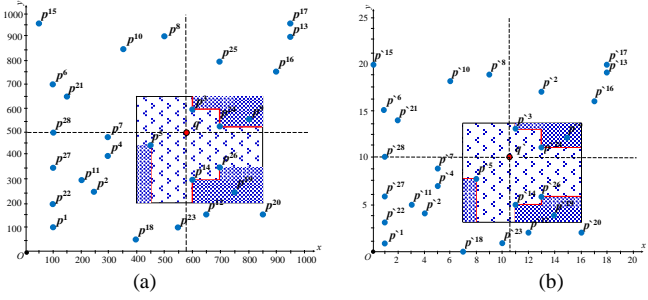

Figure 12: The constrained 1NN query with regard to a reference point $q$ lying inside the constrained region: (a) in the OS, and (b) in the ES.

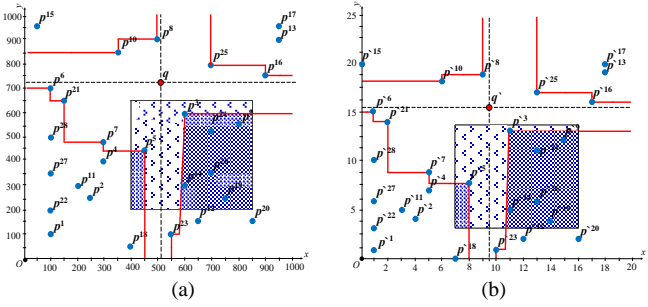

Figure 13: The constrained 1NN query with regard to a reference point $q$ lying outside the constrained region: (a) in the OS, and (b) in the ES.

### 5.8. The Constrained Skyline Query

Given a dataset $P$ of $d$-dimensional points and a set of constraints, the constrained skyline query returns the most interesting points in the data space defined by the constraints. Typically, each constraint is expressed as a range along a dimension and the conjunction of all constraints forms a hyper rectangle (referred to as the constraint region) in the $d$-dimensional attribute space **[49]**.

The SOPE model can process such queries with some necessary adaptation on the skyline query algorithm presented earlier in this paper. As in the case of the constrained $k$NN query, the entries not intersecting the constraint region are pruned.

An example of the execution of the query is presented in **Figure 14**. Based on Propositions 1 and 2 it is clearly understood that the execution of the query in both the OS and in the ES domains provide the same results.

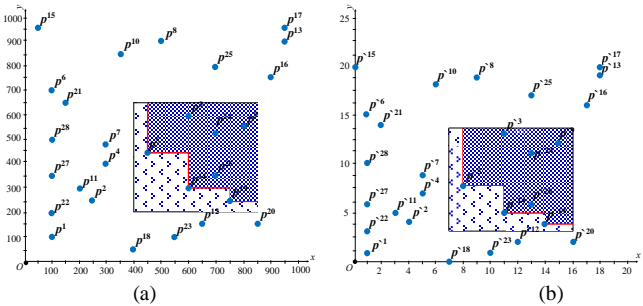

Figure 14: The constrained skyline query: (a) in the OS, and (b) in the ES.

### 5.9. The Reverse k-Nearest Neighbour Query

Given a dataset $P$ of $d$-dimensional points and a query reference point $q$, it has been seen that the nearest neighbour query retrieves the point in the dataset that is the closest to $q$. Based on this, the reverse nearest neighbour (reverse 1NN) query retrieves the points in $P$ that have $q$ as their nearest neighbour. Therefore, for the set R1NN($q$) of the reverse nearest neighbours of $q$ it holds that: R1NN($q$) = $\{p \in P: \nexists\, r \in P$ such that $distance(p, r) < distance(p, q)\}$, where $distance()$ is a specified distance metric.

The definition of the reverse 1NN query can be easily extended to the reverse $k$-nearest neighbour (reverse $k$NN) query. More specifically, the reverse $k$NN query retrieves all the points $p \in P$ that have $q$ as one of their $k$ nearest neighbours. In this case, R$k$NN($q$) = $\{p \in P: distance(p, q) \leq distance(p, r)\}$, where $r$ is the $k$-th nearest neighbour of $p\}$**[53]**. For simplicity in the following the paper will study the algorithm for supporting the reverse 1NN query and it will give some hints for its straightforward extension to cover the reverse $k$NN query as well.

For the processing of the reverse 1NN query in the proposed SOPE model the following proposition needs to be proved.

***Proposition 6:*** Let $P$ be a $d$-dimensional dataset, $q$ be the query point, GSL($q$) be the set of global skyline points of $P$ with regard to $q$ and R1NN($q$) be the set of reverse nearest neighbour points of $q$. It holds that R1NN($q$) $\subseteq$ GSL($q$).

***Proof:*** Let us suppose that we have a point $p \in$ R1NN($q$) for which $p \notin$ GSL($q$). Therefore, there is a point $r \in$ GSL($q$) that belongs in the same hyper-quadrant with $p$ with regard to $q$ which dominates $p$ in that quadrant. Therefore, since based on Definition 2, we have that $\forall\, i \in \{1, ..., d\}: (r_i - q_i)(p_i - q_i) > 0$ and $|r_i - q_i| \leq |p_i - q_i|$, and therefore $\forall\, i$ we have: $(r_i - q_i < 0$ and $p_i - q_i < 0)$ or $(r_i - q_i > 0$ and $p_i - q_i > 0)$. In the first case we then have that $r_i - q_i \geq p_i - q_i \Leftrightarrow r_i \geq p_i$, and consequently $\forall\, i$ we have: $p_i \leq r_i < q_i$. In the second case we then have that $r_i - q_i \leq p_i - q_i \Leftrightarrow r_i \leq p_i$, and consequently $\forall\, i$ we have: $q_i < r_i \leq p_i$. In both cases however we get that it is not possible that $p$ is a reverse nearest neighbour point of $q$ since between $p$ and $q$ we can find the data point $r$ in all dimensions and obviously $distance(p, r) \leq distance(p, q)$. However this is a contradiction to our hypothesis that $p \in$ R1NN($q$). Therefore $p \in$ GSL($q$) and it has been proved that R1NN($q$) $\subseteq$ GSL($q$). □

Proposition 6 proves that by calculating the GSL($q$) we can get a superset of R1NN($q$). **Algorithm 10** uses this conclusion in Line 3 in order to filter the initial dataset for the calculation of the reverse 1NN query. The server in Line 4 will then traverse the $B^+$-trees to collect the encrypted coordinates of every data point in GSL($q$) and it will send the result to the client. The client, by receiving and decrypting the GSL($q$), can execute a traditional reverse 1NN query with regard to $q$ (Line 5) in order to discard the data points in GSL($q$) which cannot be a reverse 1NN of $q$ and to consider the rest of them as candidates to be reverse 1NN points of $q$. In the final process of the algorithm in Lines 8-11, any such candidate data point $p$ will be used as a query predicate to execute a 1NN query in the encrypted dataset $P' - \{p'\}$ (Line 9) in order to retrieve its nearest point and to compare its distance from *p* with the distance between *p* and *q* (Line 10) in order to decide if *p* is indeed a reverse 1NN point of *q* (Line 11).

It needs to be reminded here that the main principle for every SOPE scheme is to not reveal any information to the server but the spatial order of the items stored. However, if by processing the reverse 1NN query, the location of the reference point *q* happens to coincides to the location of a point $r \in P$, then by informing the server (in Line 9 of the algorithm) that a point *p* is a candidate reverse 1NN of *q*, the server can infer that *r* might be the point in the dataset with the smallest distance to point *p*. For this reason, if *q* coincides to a dataset point *r*, using the Lines 6-7 of the algorithm the client will consider as candidate reverse 1NN points of *q* all the points in GSL(*q*), so that the server will inform nothing about the distance between the points in the dataset. In this case, when the client in Line 9 of the algorithm will receive the candidate nearest points of every point in GSL(*q*), s/he will automatically discard any information regarding points in GSL(*q*) for which s/he already knew from Line 5 of the algorithm that they could not be reverse 1NN points of *q*.

```
      Algorithm 10: The Reverse 1NN Query()
      Input:   a reference point q(q₁, q₂, …, q_d).
      Output:  the Reverse 1NN data points of q
 1:   Client: informs the Server to begin the
          execution of the reverse 1NN query with
          regard to the query predicate q;
 2:   Client/Server: calculate in cooperation
          the encoded version q' of q;
 3:   Server: executes the traditional global
          skyline query with regard to q', using
          the R-tree (see [50]);
 4:   Server: FOR every data point in the
          results set RS, the B⁺-tree for every
          dimension is traversed to collect the
          ciphertext of its coordinates and sends
          the point to the Client;
 5:   Client: decrypt the results and execute a
          traditional reverse 1NN query among
          them with regard to q, in order to
          select a candidate set of CS points p¹,
          p², …, p^cs that will be tested if they
          belong to the reverse 1NN set of q;
 6:   IF q belongs to the dataset P THEN
 7:      CS = RS;
 8:   FOR i = 1 to CS DO
 9:      Client/Server: execute the kNN Query(pⁱ,
          1) (Algorithm 8) by not considering pⁱ
          as a possible answer to the query;
10:      Client: IF ( distance(pⁱ, q) ≤ distance(pⁱ,
          the 1NN point of pⁱ) ) THEN
11:        Consider pⁱ in the reverse 1NN set of q;
12:   END;
```

Algorithm 10: The reverse 1NN query process in the proposed SOPE model.

The algorithm can be easily extended to cover the case of the reverse *k*NN query as well, by executing the *k*-global skyline query (instead of the global skyline query) in Line 3 and by comparing the distance between *p* and *q* in Line 10 with the distance between *p* and the *k*-th nearest neighbour of *p* in the encrypted dataset *P'* – {*p'*}.

**Figure 15** shows four different snapshots of the calculation of a reverse 1NN query example with regard to a reference point *q* in the 2-dimensional OS (**Figure 15.(a)**) and in the corresponding ES. As **Figure 15.(b)** shows, Line 3 of the algorithm will select the set of points GSL(*q*) ={$p^3$, $p^5$, $p^7$, $p^8$, $p^{10}$, $p^{14}$, $p^{21}$, $p^{23}$, $p^{24}$, $p^{26}$, $p^{28}$}. However, as **Figure 15.(c)** shows, by executing in the client-side the reverse 1NN query in the OS using as a data set the GSL(*q*), only the point $p^5$ is found to be a candidate of being a reverse 1NN of *q*, since all the other points in GSL(*q*) have at least one other point in GSL(*q*) which is closer than *q* to them. Finally, as **Figure 15.(d)** shows, in Line 7 of the algorithm, the **Algorithm 8** will be executed in the ES to find the 1NN point $p^7$ of $p^5$ which however is not closer than *q* to $p^5$, therefore $p^5$ is the reverse 1NN of *q*.

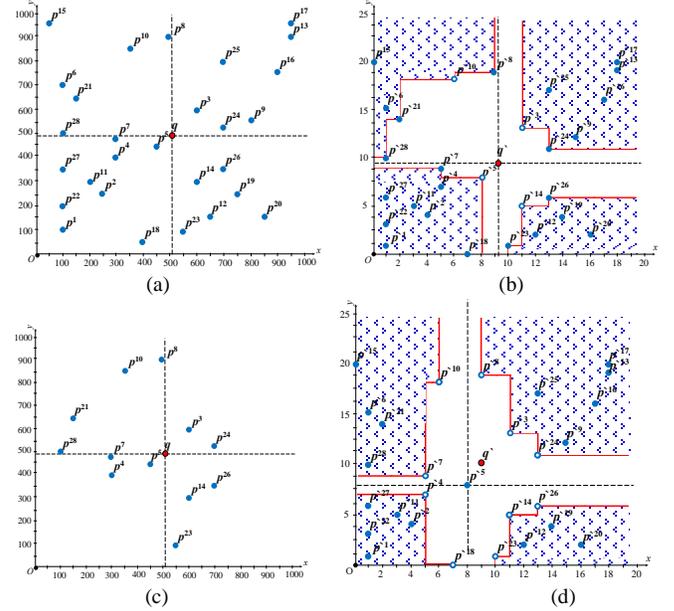

Figure 15: Four snapshots of the execution of Algorithm 10 in the example dataset for answering the reverse nearest neighbour query.

*5.10. The Continuous Nearest Neighbour Query*

Given a dataset *P* of *d*-dimensional points and a line segment *q* = [*qa*, *qb*] with *qa* and *qb* its end-points, not necessarily belonging to the dataset, the continuous nearest neighbour (continuous 1NN) query retrieves the nearest neighbour of every point in the line segment *q* (*e.g.*, "find all my nearest gas stations during my route from point *qa* to point *qb*"). In particular, the result contains a set of <*p*, [$q^1$, $q^2$]> tuples, such that the data point *p* is the 1NN of all points in the corresponding line sub-segment [$q^1$, $q^2$]. As an illustrative example **Figure 16.(a)** can be considered, in which as **Figure 16.(d)** shows the answer to the query is {<$p^{10}$, [*qa*, $q^2$]>, <$p^8$, [$q^2$, $q^3$]>, <$p^{25}$, [$q^3$, *qb*]>}, meaning that point $p^{10}$ is the 1NN for the line sub-segment [*qa*, $q^2$], *etc*. The points $q^2$ and $q^3$ of the query segment, for which we have two 1NN data points (for example the data points $p^{10}$ and $p^8$ that occur in the same distance from $q^2$ are both 1NN points of it) are known as split points **[54]**.

The definition can also be extended to the continuous *k*NN query. More specifically, the continuous *k*NN query retrieves the *k* nearest neighbours of every point in the line segment *q* (*e.g.*, "find all my 3-nearest gas stations during my route from point *qa* to point *qb*").

The algorithm for processing the continuous 1NN query in the proposed SOPE model is illustrated in **Algorithm 11**. The algorithm is adapted by a corresponding process that is introduced in **[54]** with the main difference that in the current OPE environment it needs to be executed with the cooperation of both the client and the server. Initially a SOPE-based 1NN query is performed at the two end-points *qa* and *qb* of *q* to retrieve their 1NN points (Lines 3-4). In Line 5, if *qa* and *qb* share a 1NN point then all the points of *q* share the same 1NN (this has been proved and named as *continuity property* in **[54]**). Otherwise, if *qa* and *qb* do not share a 1NN, assuming that nn*qa* is the 1NN point of *qa* and nn*qb* is the 1NN of *qb* then there is a point *qc* ∈ *q* for which the point nn*qa* is closer that nn*qb* to all the points in the segment [*qa*, *qc*] and at the same time the point nn*qb* is closer that nn*qa* to all the points in the segment [*qc*, *qb*]. This candidate spit point *qc* in *q* is calculated in Line 8 as the intersection between the query segment *q* and the perpendicular bisect-plane (which in the 2-dimensional space is converted into a perpendicular bisector) of segment [nn*qa*, nn*qb*], denoted as ⊥(nn*qa*, nn*qb*).

Since *qc* cuts up the initial line segment into two new fragments [*qa*, *qc*] and [*qc*, *qb*] the process presented in the previous paragraph is repeated retroactively for these two smaller line segments. The algorithm terminates until both the end-points $q^1$ and $q^2$ of every smaller fragment [$q^1$, $q^2$] in which the initial segment *q* has been cut up share the same 1NN point *p*. In this case, as Line 6 of the algorithm indicates, the 1NN point *p* for all the points on this segment of *q* is added in the list *SL* in the form <*p*, [$q^1$, $q^2$]>. At the end of the process the *SL* list will contain the set of tuples for the output of the continuous 1NN query with regard to *q*, however if *SL* contains continuous line segments of the form <*p*, [$q^1$, $q^2$]> and <*p*, [$q^2$, $q^3$]> that share the same 1NN point, the segments will finally be merged into the form <*p*, [$q^1$, $q^3$]>.

The algorithm can be extended to cover the case of the continuous *k*NN query as well. The reader is forwarded to **[54]** for the changes that need to be performed to the algorithm, considering that the continuity property does not hold in this case.

|   | |
|---|---|
|   | Algorithm 11: The Continuous 1NN Query()<br>Input:  a reference line segment q<br>    with end-points qa(qa₁, qa₂, …, qa_d),<br>    and qb(qb₁, qb₂, …, qb_d).<br>Output: the list SL of <p, [q¹, q²]><br>    tuples, such that the data point p<br>    is the 1NN of all points in the cor-<br>    responding line sub-segment [q¹,q²]. |
| 1: | SL = ∅; |
| 2: | Let NNqa = NNqb = ∅ be the lists of  the nearest data points to qa end   qb, respectively; |
| 3: | Client/Server: execute the kNN Que-<br>    ry(qa, 1) (Algorithm 8) and store<br>    the output in the NNqa list; |
| 4: | Client/Server: execute the kNN Que-<br>    ry(qb, 1) (Algorithm 8) and store<br>    the output in the NNqb list; |
| 5: |   |
| 6: | IF (NNqa ∩ NNqb ≠ ∅) THEN<br>    Client: Add <nnqab, [qa, qb]> tuple<br>        in list SL, where |
| 7: |     nnqab ∈ NNqa ∩ NNqb; |
| 8: | ELSE<br>    Let qc be the intersection between q<br>        and the ⊥(nnqa, nnqb), where |
| 9: |     nnqa ∈ NNqa and nnqb ∈ NNqb; |
| 10: | Client/Server: execute the Continuous<br>    1NN Query (qa, qc) (Algorithm 11); |
| 11: | Client/Server: execute the Continuous<br>    1NN Query (qc, qb) (Algorithm 11); |
| 12: | Client: IF SL holds continuous seg-<br>    ments of q that share the same 1NN<br>    point, the line segments are merged;<br>END; |

Algorithm 11: The continuous 1NN query process in the proposed SOPE model.

**Figure 16** illustrates some snapshots of an example execution of the above continuous 1NN query algorithm for a line segment *q*. For simplicity the figure shows the data only in the OS. **Figure 16.(a)** shows that in the beginning of the execution the 1NN points $p^{10}$ and $p^{25}$ of the end-points *qa* and *qb*, respectively, are retrieved from the encrypted dataset. Then, using the perpendicular bisector ⊥($p^{10}$, $p^{25}$) of segment [$p^{10}$, $p^{25}$], the candidate split point $q^1$ ∈ *q* is calculated which cuts up *q* into the line segments [*qa*, $q^1$] and [$q^1$, *qb*]. The process is now repeated retroactively for these two line segments. In the beginning of the examination of the [*qa*, $q^1$] segment, the 1NN points $p^{10}$ and $p^8$ of the end-points *qa* and $q^1$, respectively, are retrieved from the dataset. Then, as **Figure 16.(b)** shows, using the perpendicular bisector ⊥($p^{10}$, $p^8$) of segment [$p^{10}$, $p^8$], the split point $q^2$ ∈ *q* is calculated which cuts up the segment into the line segments [*qa*, $q^2$] and [$q^2$, $q^1$]. Respectively, **Figure 16(c)** shows the examination of the [$q^1$, *qb*]. In Line 5 of the algorithm, all points in the segment [*qa*, $q^2$] are discovered to share the same 1NN point $p^{10}$ and therefore the tuple <$p^{10}$, [*qa*, $q^2$]> is added to the output of the continuous 1NN query. Finally *SL* = {<$p^{10}$, [*qa*, $q^2$]>, <$p^8$, [$q^2$, $q^1$]>, <$p^8$, [$q^1$, $q^3$]>, <$p^{25}$, [$q^3$, *qb*]>} and in the final step of the algorithm (Line 11) the tuples <$p^8$, [$q^2$, $q^1$]> and <$p^8$, [$q^1$, $q^3$]> will be merged to <$p^8$, [$q^2$, $q^3$]>, as **Figure 16.(d)** shows.

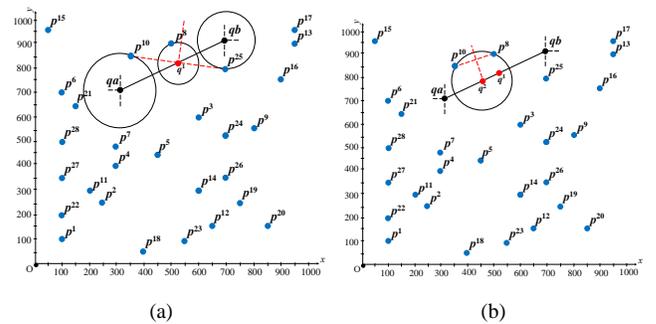

(a)          (b)

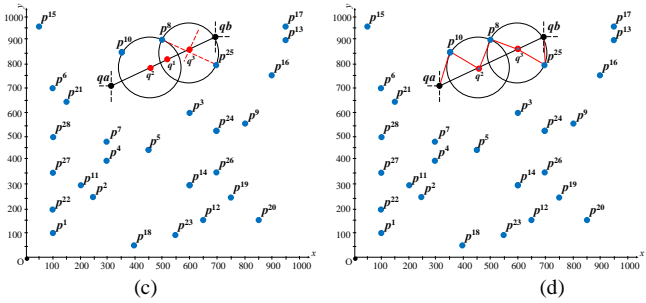

Figure 16: Snapshots of the execution of Algorithm 11 in the example dataset for answering the continuous nearest neighbour query.

## 6. Experimental study

### 6.1. Experimentation Setup

The proposed SOPE model has been implemented in Java edition 1.8 and NetBeans edition 8.1. The implementation includes the $B^+$-tree and the R-tree in their traditional form with no modification to their operations. As for the R-tree, the $R^*$-tree implementation in Java that can be downloaded from the ChoroChronos portal[2] has been utilised.

According to the proposed SOPE model, two entities participate in all the operations: the client and the server. Since only the client encrypts and decrypts the data, the symmetric key encryption has been considered for the model, although an asymmetric -public key- encryption scheme can be used as well. The particular encryption method that was used is the AES algorithm, already implemented and integrated in Java, in CBC mode and the appropriate padding (PKCS5Paddng). A 256-bits key is used which is stored on the client-side. Since Oracle Java does not support the choice of key higher than 128 out of the box, the *Java Cryptography Extension (JCE) Unlimited Strength Jurisdiction Policy Files*[3] has been employed.

The implementation of the encryption is as follows. When the client wishes to encrypt a value s/he firstly converts it to a string of characters, *i.e.*, to the string representation of this value. Then this string is encrypted using the AES algorithm with the prerequisites that were chosen earlier. The result is the so-called ciphertext. The length of the ciphertext was set to be 16 bytes.

For all experiments was used as hardware DELL laptop with Intel Core i7-4500U CPU running at 1,8GHz, with 16GB RAM and Windows 8.1 Professional 64-bit OS.

The experiments were conducted using a synthetic and a real dataset. The synthetic dataset contains 250,000 uniformly distributed 2-dimensional points that was produced by the Spatial Data Generator which can be downloaded from the Chorochronos portal. The real dataset, which was created within the AMINESS project [56], [57] and provided by the AMINESS Database [55], is also comprised by 250,000 2-dimensional points that correspond to positions of ship vessels on the Aegean Sea in several timestamps in different dates (the timestamps and the vessels' IDs were removed from the original dataset, and a set of 0.25M different 2-dimensional point locations with no duplicates was finally selected for the experimentation).

Datasets with data in higher than 2 dimensions have not been considered in this study since as it will be shown in the sequel, the performance behavior of the proposed model is expected to be burden proportionally to the number of dimensions. Using several higher-dimensional datasets for simulation, nothing unexpected which has not been already studied in the literature with regard to the performance of the $R^*$-tree and the query processing on it (such as the curse of dimensionality and the limited applicability of the skyline query in high-dimensional spaces in which most of the objects in a dataset belong to the skyline) was to be identified in our study by increasing the number of dimensions.

The disk page size was set to be 4 Kbytes and every $B^+$-tree and $R^*$-tree node is stored in a single disk page. To get a crystal-clear view on the model's performance, no buffer was used to hold any node of the indices in main memory for faster potential future usage, as for example with their most-frequently-used (MFU) or their least-recently-used (LRU) nodes. Every experiment studying the algorithms for constructing the model was repeated 10 times and at every run the objects to be inserted were shuffled and the average value of every measured parameter was calculated. An analogous strategy was followed for the query processing, some findings of which are also presented in the sequel. At every run of a query processing experiment, a different randomly selected query predicate (*i.e.*, point or/and region) was used.

In the experiments that measure the time cost performance of the proposed model for building the index and for answering queries, this cost includes also the communication cost that is spend for the interaction between the client and the server, however it does not include the cost for encrypting and decrypting the data on the client-side since the selection of the key encryption method and its performance are independent of the proposed model.

### 6.2. Data Insertion

The graph of the first experiment in **Figure 17** demonstrates the total time cost in seconds for inserting each dataset in the index, calculated as the sum of the time that is spend in the two $B^+$-trees and in the $R^*$-tree. During the data insertion process the time cost was measured at 5 intervals, *i.e.*, every 20% percent of the data being inserted. The figure shows a linear growth of the time cost as the data is inserted in the index. Due to its importance, it needs to be reminded that this time cost includes also the communication cost that is spend for the interaction between the client and the server. It is also interesting to note that in a real world application the processes in the $B^+$-trees can be performed in paralel, inependently of how many these are (*i.e.*, independently on how many the dimensions are), thus substantialy reducing the total time cost that is now illustrated.

---

[2] http://www.chorochronos.org/?q=node/43
[3] http://www.oracle.com/technetwork/java/javase/downloads/jce8-download-2133166.html

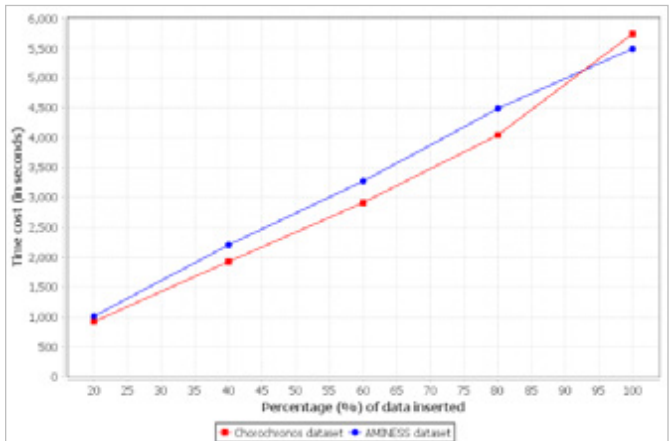

Figure 174: The total time cost for building the encrypted index, calculated as the sum of the time that is spend in the two B$^+$-trees and in the R$^*$-tree.

**Figure 18** shows the total I/O cost in disk page accesses on the server-side for the insertion of the two different datasets. As the figure shows, this cost also follows a linear increase for both datasets. Based on this experiment, **Figure 19** illustrates the average I/O cost per data object insertion and per tree (*i.e.*, the two B$^+$-trees and the R$^*$-tree) for the AMINESS dataset. The I/O cost as it is calculated for the two B$^+$-trees and for the R$^*$-tree is higher than in a traditional application with non-encrypted data, since before every data insertion the encodings of the coordinates of some other already inserted data might need to be re-calculated (by accessing some nodes on the corresponding B$^+$-trees, as it is stated in Line 5 of **Algorithm 1**) and updated (in the R$^*$-tree, as it is stated in Line 6 of **Algorithm 1**).

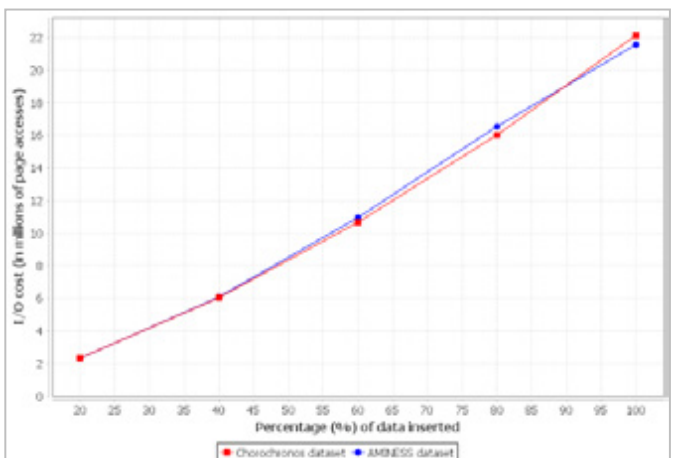

Figure 18: The total I/O cost for building the encrypted index, calculated as the sum of the I/O cost in the two B$^+$-trees and in the R$^*$-tree.

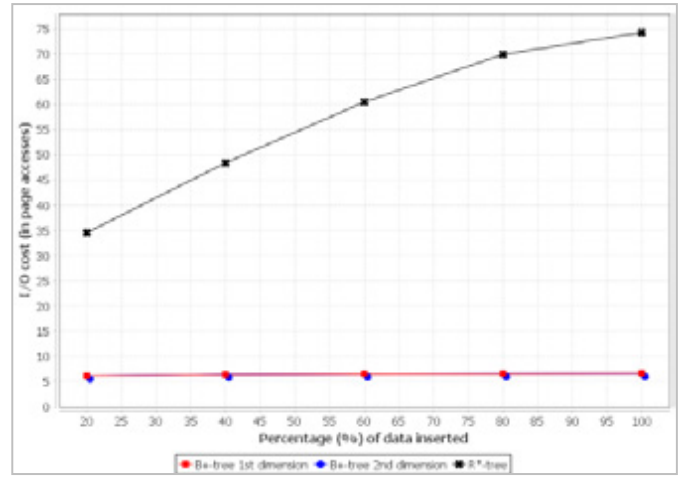

Figure 19: The average I/O cost per data object insertion for the AMINESS dataset.

The next graph in **Figure 20** studies the index size as a number of tree nodes (*i.e.*, disk pages), including the size of the two B$^+$-trees hosting the encrypted coordinates and the size of the R$^*$-tree hosting the encoded data objects. A more detailed tracing of the index size growth, seperately for the two B$^+$-trees and the R$^*$-tree for the AMINESS dataset is illustrated in **Figure 21**. The graph confirms the expected traditional $O(n/B)$ linear growth of the size of every separate index as the data are inserted. It needs to be noted that for every $d$-dimensional object the proposed SOPE model stores the ciphertexts (every one of which is 16 bytes long) of its $d$ coordinates in the corresponding B$^+$-trees and only the OPE encodings of its coordinates are stored in the R$^*$-tree.

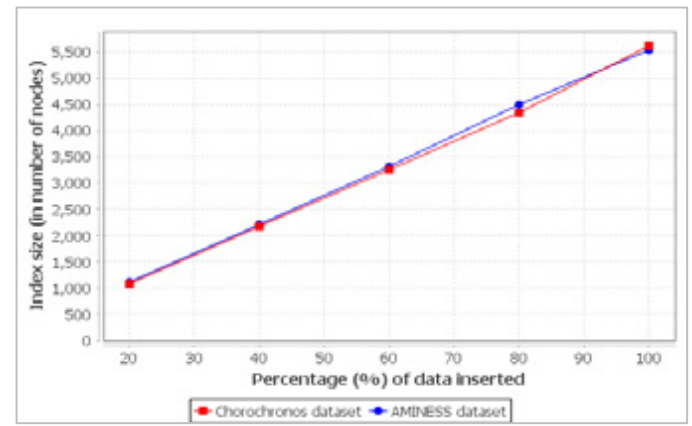

Figure 20: The size of the encrypted index, including the two B$^+$-trees and the R$^*$-tree.

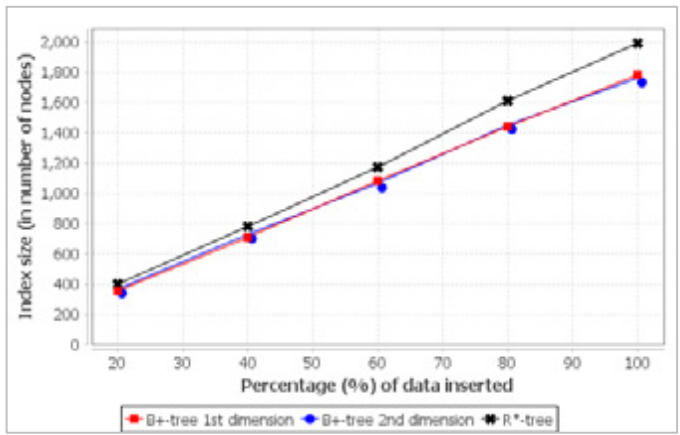

Figure 21: The size of the two B$^+$-trees and of the R$^*$-tree for the AMINESS dataset.

*6.3. Query Processing*

In the first experiment on studying the performance efficiency of the proposed model in query processing, the graph in **Figure 22** shows the time cost in milliseconds for supporting the point query. The results demonstrate that the time cost for both the real and the synthetic dataset are almost identical.

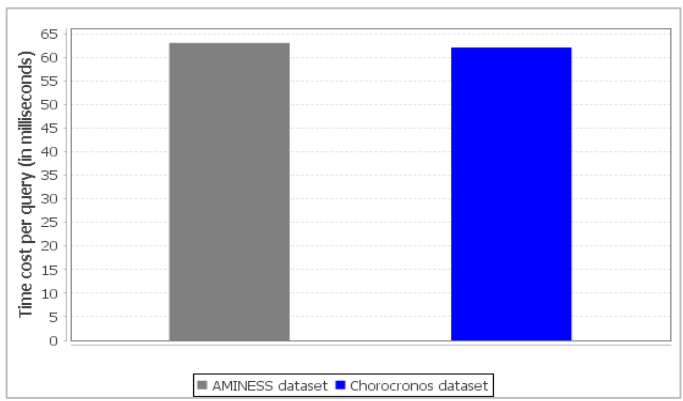

Figure 22: The time cost performance of the point query algorithm.

Next, **Figure 23**, shows the average I/O cost for answering the point query, separately for every B$^+$-tree and for the R$^*$-tree. As expected the I/O cost in that query depends only on tree height.

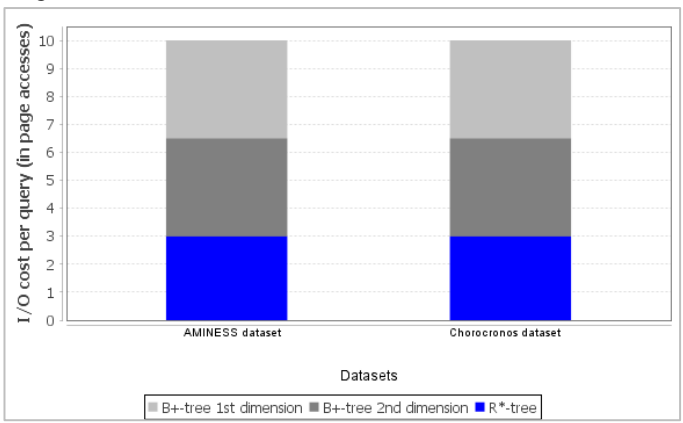

Figure 235: The I/O cost performance of the point query algorithm.

In the next experiment on studying the performance efficiency of the proposed model in query processing, the graph in **Figure 24** shows the time cost in seconds for supporting the range query with respect to a rectangular query window of three different sizes, *i.e.*, 1%, 3% and 5% of the area of the square workspace. The results demonstrate a linear growth of the time cost in relation to the window size growth. It is reminded that the time cost (for this as well as for every query that is studied in the sequel) includes also the communication cost that is spend in the steps of the **Algorithm 4** for the interaction between the client and the server.

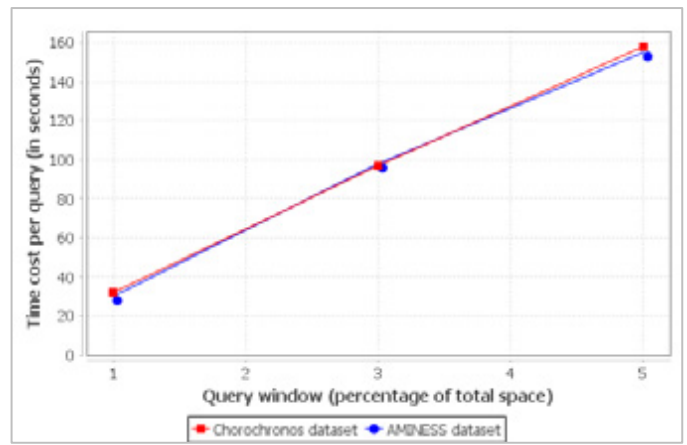

Figure 246: The time cost performance of the range query algorithm.

For the same experiment, **Figure 25** shows the I/O cost in page accesses for answering the range query, separately for every B$^+$-tree and for the R$^*$-tree for the uniformly distributed synthetic dataset (Chorocronos). The figure demonstrates that as the query window increases, so does the number of accessed nodes (*i.e.*, disk pages). To understand however the high I/O cost on accessing the B$^+$-trees, we need to focus on how the range query is processed in **Algorithm 4** and on the role that can be taken by an MFU or LRU buffer on reducing this cost, if such buffer can be provided. As the algorithm shows, every B$^+$-tree is traversed multiple times in every execution of the range query, while the R$^*$-tree is accessed only once. In Line 2 of the algorithm it is indicated that every B$^+$-tree is traversed twice to get the encoded versions of the diagonal vertices of the query hyper rectangle $q$. Also in Line 4 of the algorithm, it is indicated that every B$^+$-tree is traversed one more time for every data point in the results, in order to retrieve the corresponding ciphertexts of its coordinates. Therefore, for the uniformly distributed dataset into consideration, a query predicate $q$ covering the 1% of the workspace is expected to return about the 1% of the data points in the set, *i.e.*, 2,500 points. This means that every B$^+$-tree is traversed about 2,500 times, summing up 3*2,500 page accesses, considering a tree height of 3 pages. However, if we can keep in main memory the internal nodes of each B$^+$-tree plus one or a few of its LRU leaves (or if we can keep in main mamory just the LRU node in every level of the tree) we will greatly reduce the query processing I/O cost at least 3 times (at least 1.5 times, respectively).

It is also notable that the total time and I/O cost for processing the range query will be further reduced (more specifically, as **Figure 25** also shows, in our 2-dimensional experiments this cost will be almost halved) if the server has the power to process the two B$^+$-trees in a paraller fasion.

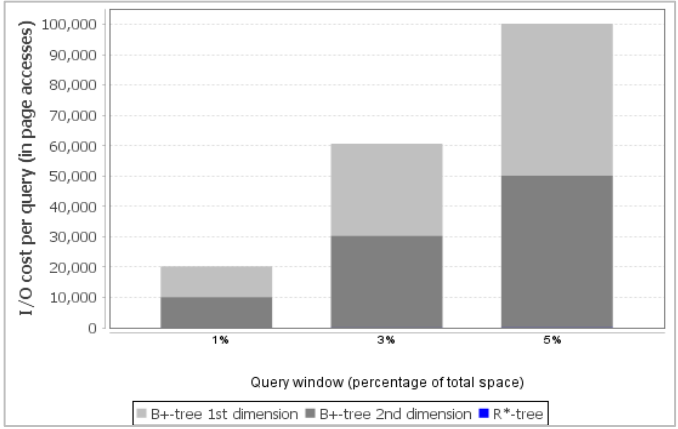

Figure 257: The I/O cost performance of the range query algorithm for the Chorocronos dataset.

The next experiment studies the performance efficiency of the (static) skyline query algorithm. **Figure 26** (**Figure 27**) demonstrates the time cost in milliseconds (the I/O cost in page accesses) for supporting the query. Both figures show a quite lower processing cost for the synthetic dataset than this of the real dataset. This result has its roots on the non-uniform distribution of the points in the real dataset that form more sparse and dense areas on the workspace than the uniformly distributed points in the synthetic dataset.

**Figure 27** shows also the much higher I/O cost for accessing the B$^+$-trees with regard to the R$^*$-tree, for the same reasons discussed in the case of the range query. As will be shown in the sequel, the same phenomenon is also apparent in the rest of the queries that are studied in this paper and the solution that was put forward in the case of the range query can be suggested to all these cases as well.

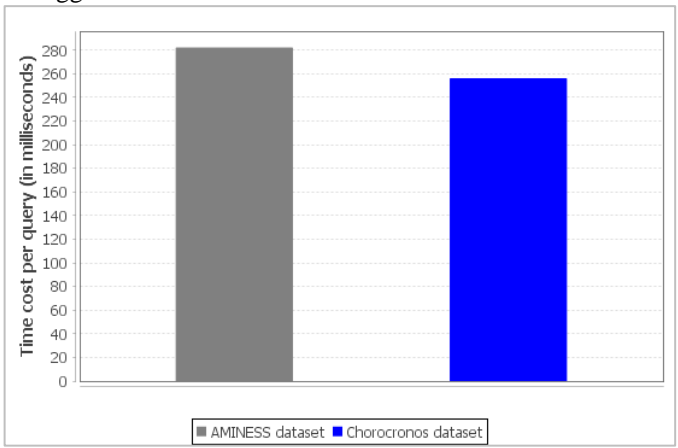

Figure 26: The time cost performance of the skyline query algorithm.

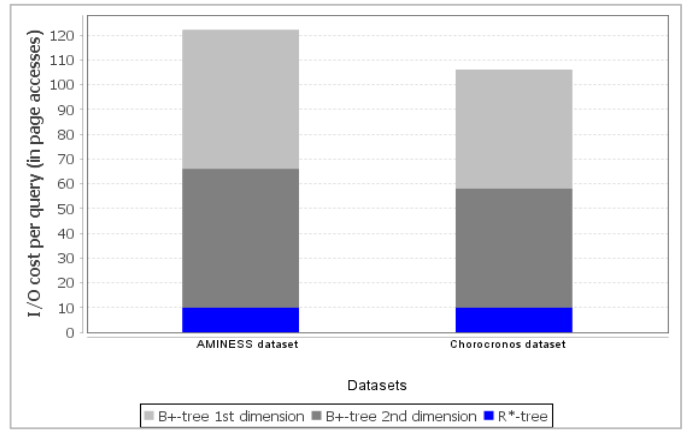

Figure 278: The I/O cost performance of the skyline query algorithm.

The next experiment studies the performance efficiency of the dynamic skyline query algorithm. **Figure 28** shows that the time cost in milliseconds for executing the query processing algorithm using the real dataset is quite smaller than that for the case of the synthetic dataset, which is the opposite conclusion than that drawn in **Figure 26** for the static skyline query. This result has its roots on the differences between the distributions of the two datasets.

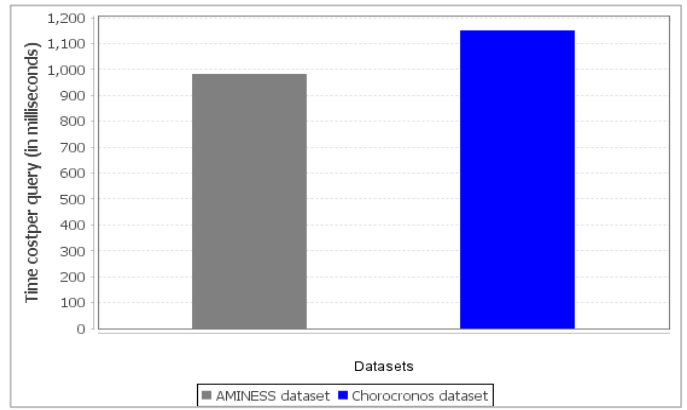

Figure 289: The time cost performance of the dynamic skyline query algorithm.

**Figure 29** demonstrates the I/O cost in page accesses for executing the dynamic skyline query algorithm as it is separately measured for every B$^+$-tree and for the R$^*$-tree.

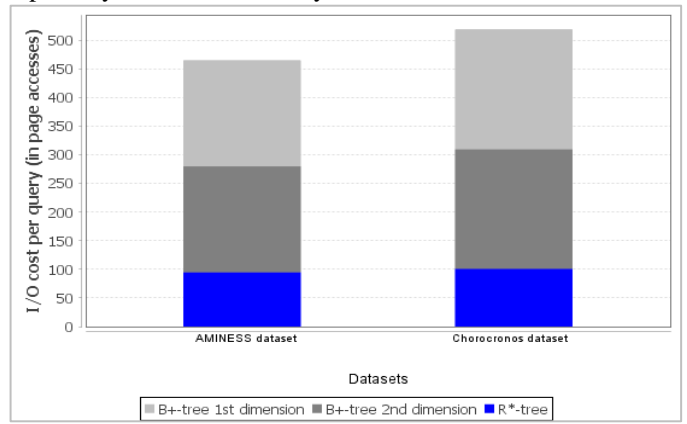

Figure 29: The I/O cost performance of the dynamic skyline query algorithm.

The next experiment studies the performance of the *k*NN query algorithm. **Figure 30** (**Figure 31**) demonstrates the time cost in milliseconds (the I/O cost in page accesses) for supporting the query for three different *k* values, *i.e.*, for *k* equal to 1, 2, and 3, for both the real and the synthetic datasets (for the synthetic dataset, respectively). As expected the linear growth of the parameter *k* increases linearly the cost for processing the query.

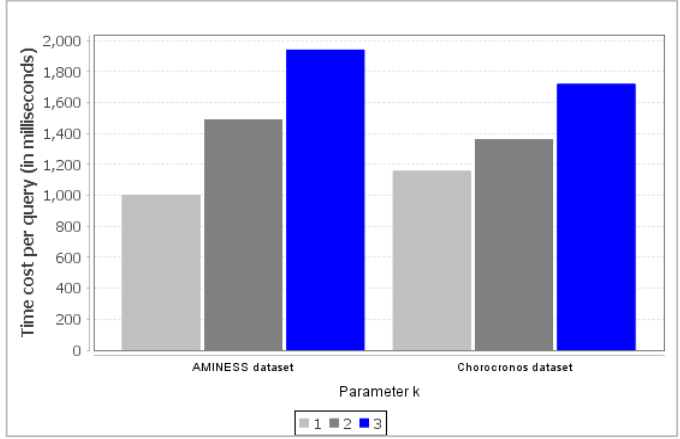

Figure 30: The impact of the parameter *k* on the time cost performance of the *k*NN query algorithm.

**Figure 31** shows also the I/O cost separately for every $B^+$-tree and for the $R^*$-tree. An interesting conclusion is that the number of disk page accesses in the $R^*$-tree does not increase much from one *k* value to the next. This happens because retrieving the *k*-th and the *k*+1-th global skyline often means retrieving spatially adjacent data points which are stored on the same $R^*$-tree nodes. Thus the number of pages for retrieving the first *k* global skyline sets or the first *k*+1 global skyline sets does not differ much when -as **Algorithm 12** in the Appendix shows- this retrieval operation is performed in a single $R^*$-tree traversal.

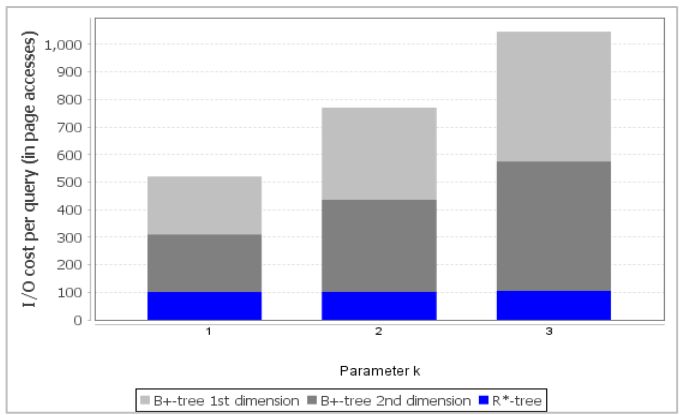

Figure 31: The impact of the parameter *k* on the I/O cost performance of the *k*NN query algorithm for the ChoroChronos dataset.

The next experiment investigates the performance of the proposed model in processing the constrained *k*NN query. **Figure 32** (**Figure 33**) demonstrates the time cost in milliseconds (the I/O cost in page accesses) for supporting the query for the ChoroChronos synthetic dataset, using three different values for the parameter *k* and two different sizes for the constrained area. The figures again show the proportional growth of the query processing cost with respect to these two input parameters. By comparing the results in **Figures 32** and **33** with the corresponding results in **Figures 30** and **31** for the (un-constrained) *k*NN query algorithm, we can apprehend the time and I/O costs reduction that is introduced by the computation of the *k*NN query in a constrained region. It is notable that with regard to the $R^*$-tree this cost reduction cannot be exactly proportional to the percentage of the workspace that the constrained region represents due to the overlapping of this region with the MBRs of a higher percentage of tree nodes.

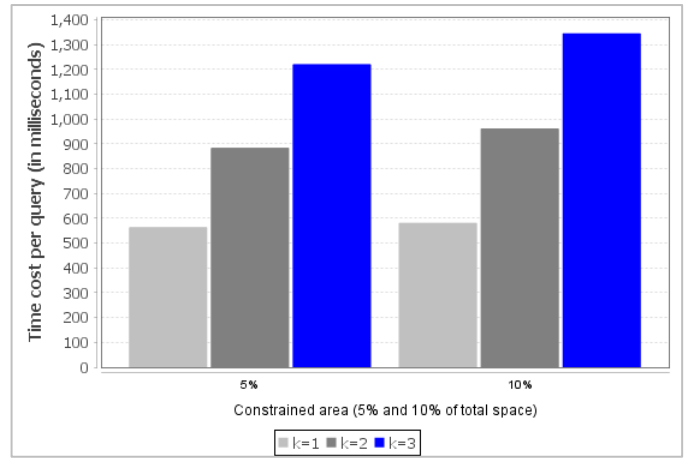

Figure 32: The impact of the constrained area size and of the parameter *k* on the time cost performance of the constrained *k*NN query algorithm for the ChoroChronos dataset.

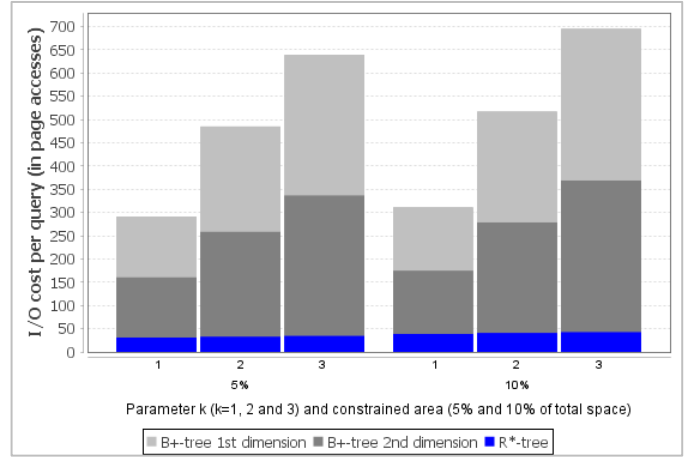

Figure 33: The impact of constrained area size and of the parameter *k* on the I/O cost performance of the constrained *k*NN query algorithm for the ChoroChronos dataset.

The study on the performance efficiency of the constrained skyline query algorithm is omitted because it draws similar conclusions to the ones in the case of the constrained *k*NN query algorithm.

The next experiment studies the performance efficiency of the reverse *k*NN query algorithm. **Figure 34** (**Figure 35**) demonstrates the time cost in milliseconds (the I/O cost in page accesses) for supporting the query for three different *k* values, *i.e.*, for *k* equal to 1, 2, and 3, for both the real and the synthetic datasets (for the synthetic dataset). As expected, the linear growth of the parameter *k* increases linearly the cost for processing the query. Another interesting conclusion that is drawn by comparing the **Figures 34** and **35** to the **Figures 30** and **31** is that the execution time and I/O costs when processing the reverse $k$NN query are about 4 times higher than when processing the $k$NN query for the same datasets. The root of this difference is in the Lines 8-11 in the reverse $k$NN query algorithm (**Algorithm 10**) that do not exist in the $k$NN query algorithm (**Algorithm 8**). The 'FOR' loop in these lines of the algorithm executes the $k$NN query algorithm for every candidate reverse $k$NN of $q$, and the findings show that on average 3 such candidate points exist in every case. This findings come in harmony with **[58]** which comes to the conclusion that for $k = 1$ in the 2-dimensional space there are 6 candidate reverse nearest neighbour points at the most to any query predicate point $q$. Therefore the Lines 8-11 in the reverse 1NN query algorithm cannot be executed more than 6 times at the most (it needs to be noted here that in our study no reference point $q$ happened to coincide to the location of any point $r$ in the dataset).

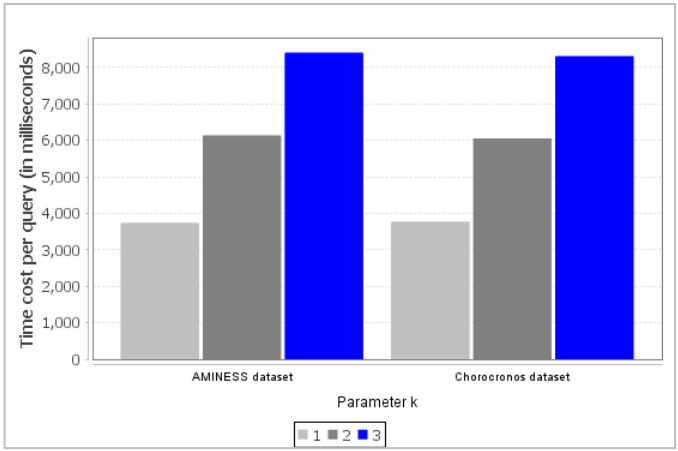

Figure 3410: The impact of the parameter $k$ on the time cost performance of the reverse $k$NN query algorithm.

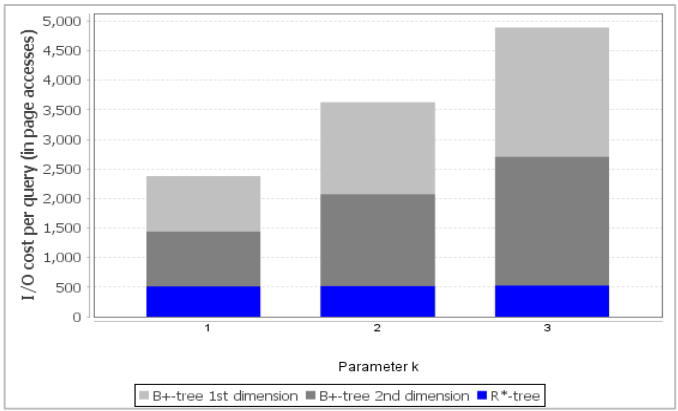

Figure 35: The impact of the parameter $k$ on the I/O cost performance of the reverse $k$NN query algorithm for the ChoroChronos dataset.

The final experiment studies the performance efficiency of the continuous 1NN query algorithm. **Figure 36** (**Figure 37**) demonstrates the time cost in milliseconds (the I/O cost in page accesses) for supporting the query using three different lengths of the query line, *i.e.*, 1%, 3% and 5% of the length of the side of the square workspace. The figures again show the proportional growth of the query processing cost with respect to the query length.

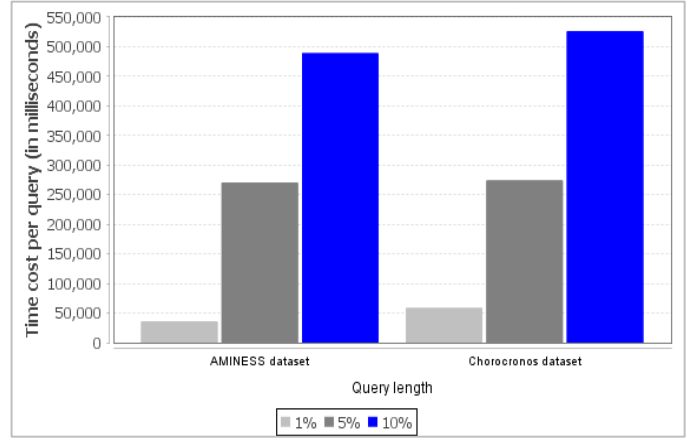

Figure 36: The impact of query length on the time cost performance of the continuous 1NN query algorithm.

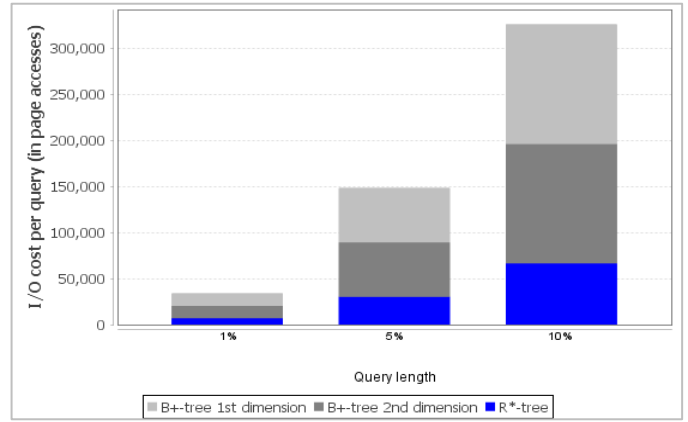

Figure 37: The impact of line length on the I/O cost performance of the continuous nearest neighbour query algorithm for the ChoroChronos dataset.

## 7. Conclusion

The paper introduces the Spatially-based Order-Preserving Encryption (SOPE) model for $d$-dimensional data. The new model is inspired and extends the well-known OPE model in $d$-dimensional databases; therefore, it supports the safe storage and efficient retrieval of $d$-dimensional data to a remote untrusted server by revealing to the server nothing else but the spatial order of the data objects.

The paper proposes algorithms for constructing the model and for efficiently processing a large set of 10 or more popular queries for $d$-dimensional data, such as the point query, the range query, the (static, dynamic, and global) skyline query, the $k$-nearest neighbour query, the reverse $k$-nearest neighbour query, the continuous nearest neighbour query, *etc*. **[59]**.

The prototype implementation and experimental evaluation of the proposed model based on synthetic and real data indicates that the new model is able to provide searchable encryption for $d$-dimensional cloud databases in modern applications in which there is no need to keep secret the corresponding spatial relations between these objects.

As regard to future plans of research, the building and querying operations algorithms of the proposed model will need to be examined for possible optimizations, towards improving the speed of the efficiency of the model. Such optimizations might be to use buffers in the trees when executing the query processing algorithms, or to store the ciphertexts of the coordinates of the data objects directly into the R-tree instead of the corresponding B$^+$-trees, so that the server will not need to traverse every B$^+$-tree to retrieve these ciphertexts before s/he will return the output of a query to the user.

Also, although several other known spatial queries can be easily supported by the proposed model (for example the reverse skyline query can be easily implemented using the global query processing **Algorithm 6** proposed in this paper in combination with heuristics proposed in **[58, 50]** that are still valid under the SOPE model), there are spatial queries that need to be further studied (for example the closest-pair query **[60]**, *etc*.).

**Acknowledgments**

We would like to thank Mr. Kalyvas Christos (chkalyvas@aegean.gr) and Mr. Kokkos Athanasios (ath.kokkos@aegean.gr), the AMINESS Database administrators / creators, for providing us the AMINESS dataset.

## Appendix

The algorithm for processing the $k$-global skyline query [50] using an R-tree, which is based on the traditional (1-)global skyline algorithm that is described in **Algorithm 6**, follows in **Algorithm 12**. For the sake of completeness, the formal presentation of the supporting *GloballyDominated* function for finding if a $d$-dimensional object is globally dominated by some point from a set of points is presented in **Algorithm 13**.

```
       Algorithm 12: The k-Global Skyline
          Query()
       Input:  a reference point q(q₁, q₂, …, q_d)
          and the requested value of k.
       Output: a list GSL[0..k-1] of data points
          organised in k different sets,   every
          one of which represents the correspon-
          ding i-th global skyline with regard
          to q, ∀ i ∈ {1, …, k}.
1:     GSL[0..k-1] = ∅;
2:     Let H = ∅ be a heap of records;
3:     Insert all entries of the root of the
          R-tree in H and sort the entries in H
          in an ascending order according to
          their distance from q;
4:     WHILE H is not empty DO
5:        Remove the first entry e of H for
             examination;
6:        IF GloballyDominated(e, GSL[0..k-1],q)
             THEN
          //i.e., if e is k-globally dominated
          // by some point in GSL w.r.t. q
7:           Discard e;
8:        ELSE IF e is an intermediate entry THEN
9:           FOR each child record eⁱ of e DO
10:             IF !GloballyDominated(eⁱ,
                   GSL[0..k-1], q) THEN
             //i.e.,if eⁱ is not k-globally domi-
             //nated by some point in GSL w.r.t. q
11:                Insert eⁱ into H by keeping H
                      sorted;
12:       ELSE
13:          FOR i=0 TO i=k-1 DO
14:             IF !GloballyDominated(e,GSL[i],q)
                   THEN
15:                Insert e into GSL[i]; Break;
16:    return GSL[];
```

```
       Algorithm 13: The GloballyDominated
          Function()
       Input:  a query point q(q₁, q₂, …, q_d) and
          a d-dimensional point e(e₁, e₂, …,
          e_d) or a hyper rectangle er
          with its vertex e that is closer to q
          having coordinates e(e₁, e₂, …, e_d),
          a list GSL() of data points.
       Output: TRUE if e, or er, respectively, is
          globally dominated by some point in
          GSL w.r.t. q, and FALSE otherwise.
1:     Integer first, second; Boolean third;
2:     FOR every data entry i in GSL() list DO
3:        Set first = second = 0; third = false;
4:        FOR every dimension j DO
5:           IF (GSL(i)_j - q_j)(e_j - q_j)> 0 THEN
6:              first++;
7:           IF |GSL(i)_j - q_j| ≤ |e_j - q_j| THEN
8:              second++
9:           IF |GSL(i)_j - q_j| < |e_j - q_j| THEN
10:             third = true;
11:       IF (first == d) AND (second = d)
             AND third THEN
12:          return TRUE;
13:    return FALSE;
```

The algorithm for processing the constrained $k$-global skyline query using an R-tree follows.

```
       Algorithm 14: The Constrained k-Global
          Skyline Query()
       Input: a reference d-dimensional hyper
          rectangle r with diagonal vertices
          ra(ra₁, …, ra_d), and rb(rb₁, …, rb_d),
          a query point q(q₁, …, q_d) and the
          requested value of k.
       Output:  a list GSL[0..k-1] of data points
          organised in k different sets,   every
          one of which represents the correspon-
          ding constrained i-th global skyline
          with regard to q, ∀ i ∈ {1, …, k}.
1:     GSL[0..k-1] = ∅;
2:     Let H = ∅ be a heap of records;
3:     Insert in H all the entries of the root
          of the R-tree that intersect the   region r
       and sort the entries in H in
          an ascending order according to their
          distance from q;
4:     WHILE H is not empty DO
5:        Remove the first entry e of H for
             examination;
6:        IF GloballyDominated(e, GSL[0..k-1], q)
             THEN
          //i.e., if e is k-globally dominated
          // by some point in GSL w.r.t. q
7:           Discard e;
8:        ELSE IF e is an intermediate entry THEN
9:           FOR each child record eⁱ of e DO
10:             IF ((eⁱ intersects r) AND
                   !GloballyDominated(eⁱ,
                      GSL[0..k-1], q)) THEN
             //i.e.,if eⁱ is not k-globally domi-
             //nated by some point in GSL w.r.t. q
11:                Insert eⁱ into H by keeping H
                      sorted;
12:       ELSE
13:          FOR i=0 TO i=k-1 DO
14:             IF !GloballyDominated(e, GSL[i], q)
                   THEN
15:                Insert e into GSL[i]; Break;
16:    return GSL[];
```